# Gender income disparity in the USA: analysis and dynamic modelling

Ivan O. Kitov[1] and Oleg I. Kitov[2]

[1] The Institute of Geospheres Dynamics, Russian Academy of Sciences
[2] The University of Oxford

Abstract
We analyze and develop a quantitative model describing the evolution of personal income distribution (PID) for males and females in the U.S. between 1930 and 2014. The overall microeconomic model, which we introduced ten years ago, accurately predicts the change in mean income as a function of age as well as the dependence on age of the portion of people distributed according to the Pareto law. As a result, we have precisely described the change in Gini ratio since the start of income measurements in 1947. The overall population consists of two genders, however, which have different income distributions. The difference between incomes earned by male and female population has been experiencing dramatic changes over time. Here, we model the internal dynamics of men's and women's PIDs separately and then describe their relative contribution to the overall PID. Our original model is refined to match all principal gender-dependent observations. We found that women in the U.S. are deprived of higher job positions (work capital). This is the cause of the long term income inequality between males and females in the U.S. It is unjust to women and has a negative effect on real economic growth. Women have been catching up since the 1960s and that improves the performance of the U.S. economy. It will take decades, however, to full income equality between genders. There are no new defining parameters included in the model except the critical age, when people start to lose their incomes, was split into two critical ages for low-middle incomes and the highest incomes, which obey a power law distribution. Such an extension becomes necessary in order to match the observation that the female population in the earlier 1960s was practically not represented in the highest incomes. In the overall model, male population dominate in the top income range and the difference between two critical ages is masked. Gender versions of the refined model provide consistent quantitative description of the principal features in the male and female income distribution.

Key words: income, gender, age, USA, modelling

JEL Classification: D01, D31, E17, E64, J1, O12

## Introduction

Income distribution is a multidimensional socio-economic process observed, studied, controlled, and reported by a multitude of organization from governmental statistical and research agencies to universities and private companies. The most basic measurements are carried out at the level of personal income, which is measured in domestic currency. All individual incomes are always quantitatively defined in the same monetary units whatever is the form and source of income. The result of income measurements is expressed in numbers and available for the brooder research community as tables and time series suitable for quantitative description and prediction. In the current social and political discussions of economic inequality, household (family) income is a more frequent guest, however. The household income aggregates the measured personal figures according to changing definition of a household and is subject to secular and nonlinear variations in composition. As it is a derivative measure of income, which is prone to higher fluctuations over time, the quantitative analysis and modelling have to address the dynamics of indivisible entities – personal incomes.

The process of income distribution is an intrinsically dynamic one and has clear features changing with time, but it also demonstrates a higher stability of aggregate measures of inequality. For example, the Gini coefficient measured by the Census Bureau for personal incomes has not been changing since the 1960s, as prove the reports of the CPS (Current Population Survey) Annual Social and Economic (ASEC) Supplements [U.S. Census Bureau, 2015a]. If the Gini ratio is the same from 1960 to 2014, the underlying Lorentz curve implies that in any given range of total income there is always the same portion of total population.

The income inequality is fixed with just small variations induced by the change in age pyramid.

Unequal division of incomes between people raises many research problems, which include gender disparity [*e.g.,* U.S. Census Bureau, 2000**;** Sumati, 2007; Shaikh *et al*., 2007], age dependence [Mincer, 1958, 1974; Hartog, 2004; Kitov 2006], poverty [*e.g.,* Jenkins, 2007; DeNavas-Walt and Proctor, 2015], the mechanisms of distribution [*e.g*., Galbraight, 1998; Neal and Rosen, 2000; Kitov, 2005a; Jenkins, 2009], top incomes [*e.g*., Piketty and Saez, 2003; Jenkins *et al*., 2009; Auten and Gee, 2009; Auten *et al*., 2013] national/international similarities and differences [*e.g*., Atkinson *et al*., 1995; Gottschalk and Smeeding, 1997; Atkinson and Brandolini, 2001; Bandourian *et al*., 2003] among many others. It is important to have a numerical model merging all quantitatively defined aspects of income distribution by a parsimonious set of dynamic equations, which would serve as a workhorse to the broader research community, policy making bodies, and governmental services.

We developed a microeconomic model in the mid-2000s and demonstrated that the dynamics of personal income distribution (PID) observed in the USA since 1947 (the start of personal income measurements) can be accurately described quantitatively [Kitov, 2005a; Kitov and Kitov, 2013]. Our model (further KKM) predicts the age dependence and the evolution with time of the overall PID and all its derivatives like mean income and the portion of people in the Pareto distribution. These features demonstrate significant changes over time and reveal measureable sensitivity to age. The KKM well predicts the larger changes in the PID and marginal changes in the Gini coefficient as observed since 1947.

The age dependence of personal income distribution is a well-known dimension of income inequality. Age is a natural parameter causing unequal income sharing. Everyone has to start and end with zero income. Between two zeroes it should be a maximum. The extent of age-driven inequality is a different issue. There are ongoing complaints that, in relative terms, younger people are getting poorer and poorer over time. The KKM describes the age dependence and its evolution as a function of real GDP per capita.

As an exogenous and independent variable, the GDP time series is fixed for modelling purposes. The only improvement we have introduced in the original real GDP per capita estimates reported by the Bureau of Economic Analysis [U.S. BEA, 2015] is a correction for the difference between total and working age population. Between 1947 and 2014, the ratio of the total and working age population varies from 1.45 in 1961 to 1.26 in 2014. Since income is defined only for people of 15 years of age and above the use of the original GDP would introduce a significant bias in the model. The change in the corrected real GDP is much smaller than in the BEA's time series. Such a correction is a must for cross country comparison of personal income distribution [Kitov and Kitov, 2015].

Gender is another dimension of income inequality, which also reveals the age-dependence. Male and female population differs in total income as well as in the portion of people with income as defined by the Census Bureau. In this study, we describe the difference in income distribution between men and women in the U.S. using internal parameters of our model and explain the evolution of this difference as a function of real GDP per capita. We predict two PIDs and their evolution independently.

Quantitative description of income distribution as driven by only one measurable economic force is a challenge for many economic models involving personal qualities or human capital, stochastic processes, selection mechanisms, etc. When a numerical model is correct it predicts each and every income depending on age, gender, race and other characteristics but does not personalize them. Standard economic consideration of individual choice in varying configurations of external and internal conditions is likely irrelevant because when all choices are aggregated they have no net effect on the final distribution of personal incomes. All individual choices and efforts, all possible interactions between people and so on end up in an accurately predicted number of people with a given gender-race-age kit having incomes within predefined range.

The model anonymizes people to the extent their qualities and choices do not matter for income distribution except those characterizing their positions in the model. In reality, people are not anonymous and their lives are not predefined at all. However, their choices are rational to the extent they are controlled by interactions with other people and nature, i.e. by the stiff socio-economic structure. As a result, the relative shape of PID, and thus, the Gini coefficient for personal incomes measured by the Census Bureau do not change with time. There are strong internal dynamic processes related to age and gender, which are analyzed and modelled in this paper. The effects associated with race are studied in a separate paper. A cross-country comparison based on income data from the U.S., UK, Canada, and New Zealand reveals the almost identical features of personal income distribution, but observed with time delays defined by the difference in real GDP per capita [Kitov and Kitov, 2015].

We start with a comprehensive description of our microeconomic model illustrated by comparison of predicted and measured features. To better characterize the evolution of the overall PID we match the dynamics of the age-dependent mean income and the portion of people in the Pareto distribution, i.e. those people whose incomes are distributed by a power law. In Section 2, we present selected features of the measured income inequality to highlight and quantify the difference between genders. At this stage, the observed differences are qualitatively interpreted in terms of model-related parameters, *i.e.* we discuss possible forces behind these differences. Section 3 is devoted to modelling. All KKM defining parameters are calibrated to the observed features of income distribution. The parameters providing the best fit are then discussed in terms of reasons for gender income inequality and their socio-economic consequences.

1. **Microeconomic model**
   1.1. **Physical intuition behind income growth and fall**

Here, a microeconomic model is presented, which has been developed to quantitatively describe the dynamics of personal income growth and distribution [Kitov, 2005a]. The model is based on one principal assumption that each and every individual above fifteen years of age has a personal capability to work. In essence, the capability to work is equivalent to the capability to earn money. To get money income, individuals have to use one or several means or tools from the full set of options that may include paid job, government transfers, bank interest, capital gain, inter-family transfers, and others. The U.S. Census Bureau questionnaire [2006] lists tens of money income components. It is important to stress that some principle sources of income are not included in the CB definition, which results in the observed discrepancy between aggregate (gross) personal income (GPI), as reported by the Bureau of Economic Analysis and the gross money income calculated by the CB.

In this section, we summarize the formulation of a theoretical model, originally described in Kitov [2005a], and present it as a closed-form solution in a simplified setting. Figure 1 illustrates a few general features any consistent model has to describe quantitatively. In the left panel, we display the evolution of mean income curves from 1962 to 2013. The original income data are borrowed from the Integrated Public Use Microdata Series (IPUMS) preparing and distributing data for the broader research community [King *et al*., 2010]. These are income microdata, *i.e.* each and every person from the IPUMS tables is characterized (among other features) by age, gender, race, gross income, and the population weight, which allows projection the individuals from the CPS population universe to the entire population. Using age, income, and population weight we have calculated the age dependent mean income for all years and then normalized them to their respective peak values. The normalized curves better illustrate the growth in the age of peak income – from below 40 in the earlier 1960s to 55 in the 2010s. This is a sizeable change likely expressing the work of inherent mechanisms driving the evolution of personal income distribution. One cannot neglect the effect of increasing age when people reach their peak incomes – neither from theoretical nor from the practical point of view.

In the right panel of Figure 1, we compare various mean income curves reported by two different organizations responsible for income measurements: the Census Bureau (CPS) and the Internal Revenue Service (IRS). The latter organization does not publish the age distribution of income at a regular basis and only the year of 1998 is available for such a comparison. The IRS mean income is calculated in 5-year age cells [IRS, 2015], the CPS prepares historical datasets with a 5-year granularity since 1993, and the annual estimates are available from the IPUMS microdata. The annual curve has also been smoothed with a nine-year moving average, MA(9). As in the left panel, all curves are normalized to their peak values.

There are significant differences in income sources and population coverage used by the CPS and IRS [Kitov, 2014]. Nevertheless, between 40 and 60 years of age, all curves in the right panel of Figure 1 are close to each other. With regard to the age of peak income, the CPS and IRS give identical results to the extent the age aggregation allows. The IPUMS curve has been smoothed and thus might have a slightly biased peak age. Between 25 and 40 years of age, the difference in normalized mean income is larger - likely because of the difference in income sources. Same effect is observed in the eldest age groups, where taxable incomes are not so often and the CPS curve is above the IRS one.

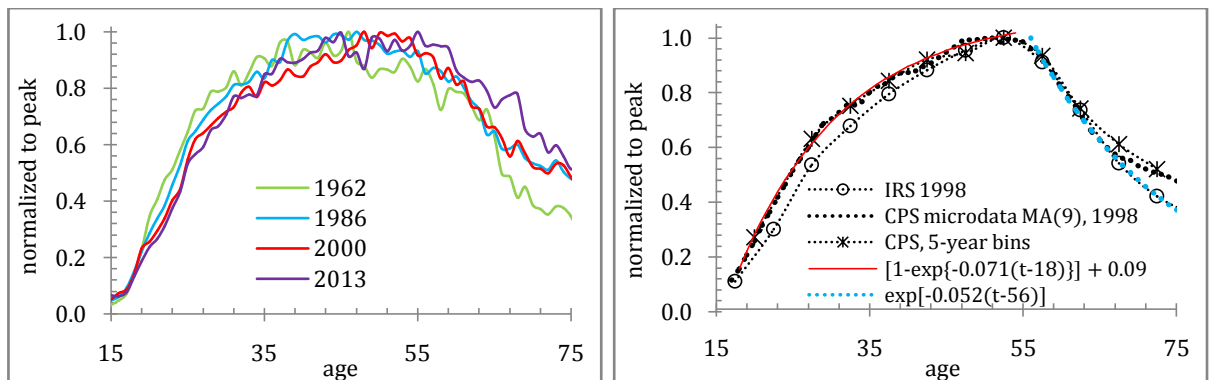

Figure 1. Left panel: The change in the shape of mean income dependence on age from 1962 to 2013 as measured by the Census Bureau in the March Supplements of the Current Population Survey. All curves are normalized to their respective peak values. Right panel: Comparison of mean income dependence on age as measured by the Census Bureau (CPS) and the Internal Revenue Service (IRS). The only year with data available from the IRS is 1998. Red line – function approximating the IPUMS curve between 18 and 55 years of age. Blue line - function approximating the IRS curve above 56 years of age.

The closeness of the peak ages measured by the IRS and CPS is important for model applicability and reliability. The accuracy of income measurements, the coverage of population and income source, the level of historical consistency in income definition and survey methodology, the entire diversity of personal characteristics, and the length of time series provided by the Census Bureau all these features make it inevitable to use the CPS data for quantitative modelling. The reverse side of this choice is the necessity to defend the modelling results against the accusation that the CPS data are not full and representative.

It is true that the CPS misses some important sources of higher incomes, but Figure 1 stresses that the estimates of key features are not different if the IRS sources are included and some CPS income sources are excluded [Henry and Day, 2015, Ruser *et al.*, 2004; Weinberg, 2004; U.S. Census Bureau, 2015b]. Besides, the CB provides the best income estimates for the poorest population, where incomes are just several dollars per year. Other organizations ignore small incomes. As a result, the estimates of personal income inequality based on the IRS data exclude half of population, the poorest half. It is difficult to consider such estimates as accurate and helpful for understanding the mechanisms of income distribution. The BEA income data are worthless for quantitative analysis of individual incomes - no age, gender, race information is available.

Astoundingly, the principal features observed in Figure 1 can be accurately approximated by basic mathematical functions. Moreover, these functions represent solutions of simple ordinary differential equations. Solid red line in the right panel is calculated to fit the CPS mean income curve. For this line, the equation is [1 - *exp*(-0.071(*t*-18))] + 0.09, where *t* is the age. The overall fit between the measured and approximating curves is extremely good from 18 to 55 years of age, before the mean income curve starts to fall.

The approximating equation is a well-known function often called "exponential saturation function". This function represents a closed-form solution of a simple ordinary differential equation d$x$(*t*)/d*t*=*a*-*b*x(*t*), where *a*>0 and *b*>0 are constants. The match between the observed and approximating curves provides some hint on the forces behind income growth. Second term in the above equation represents the force counteracting the unlimited growth of x(*t*). The amplitude of the counteracting force is proportional to the attained level, and that implies the finite value of x(*t*)<$X_{max}$, *t* →∞.

A standard example in general physics to illustrate the saturation process is associated with heating of a metal ball by an internal source with constant power, U. The growth in temperature, *T*, is balanced by energy loss through the surface, and the energy flux through the surface is proportional to the attained temperature. Thermal conductivity can be treated as infinite in terms of the characteristic time of all other processes. For a ball of radius *R* and volumetric heat capacity, $C_v$, one can write the following equation:

$$4/3\pi R^3 C_v dT(t)/dt = U - DT(t)4\pi R^2 \qquad (1)$$

where *D* is a constant defining the efficiency of heat loss through the surface, which is similar to dissipation. By dividing both sides of (1) by $4/3\pi R^3 C_v$ we obtain:

$$dT(t)/dt = \tilde{U} - \tilde{D}T(t)/R \qquad (2)$$

where $\tilde{U}=3U/(C_v 4\pi R^3)$ is the specific power of the heating sources expressed in units of thermal capacity, and $\tilde{D} = 3D/C_v$. The solution of (2) is as follows:

$$T(t) = T_0 + (\tilde{U}R/\tilde{D})[1 - exp(-\tilde{D}t/R)] \qquad (3)$$

Relationship (3) implies that temperature approaches its maximum value $\tilde{U}R/\tilde{D}$ along the saturation trajectory, which we also observe in Figure 1. Instructively, the maximum possible temperature is proportional to *R*. This fact is helpful and important for better understanding of our model and income observations. We interpret temperature as income, which one can reach using some physical capital, say, $4/3\pi R^3$, and personal efforts, say, U. Then the saturation curve in Figure 1 becomes an obvious result.

Above the age of peak mean income in Figure 1, one observes an exponential fall. Blue dotted line is defined by function *exp*[-0.052(*t*-56)]. It best matches the IRS curve above 56 years of age. The match between the observed curve and the exponent is extraordinary even in terms of the hard sciences. The exponential function is a solution of a familiar equation:

d$x(t)$/d$t$=-$b$x($t$). The only difference is in the absence of term *a*, but now the curve starts from 1.0. The evolution of mean income measured by the IRS above the critical age can be expressed by a differential equation formally identical to that describing free cooling of a preheated sphere, *i.e.* when heating source U=0 in (1).

Hence, the observed features of the mean income behaviour are similar to those observed in simple physical experiments. However, we need to describe income trajectory for each and every person in a given economy. It is natural to suggest that all individual incomes follow own saturation curves and their average value follows up some individual trajectory. Then the distribution of parameters defining individual trajectories, *i.e.* income analogues of R and U, is completely constrained by observations. This is the intuition behind our microeconomic model.

Originally, the idea of income modelling with equation (2) came from geomechanics [Rodionov *et al.*, 1982]. An identical equation describes the growth of stress, σ($t$), in an inhomogeneous inclusion with characteristic size *L* experiencing deformation at a constant rate $\dot{\varepsilon}$ as induced by external forces. Solution (3) is important to predict the highest possible level of stress at a given inclusion with size *L*. Unlike in the simple experiment with heated sphere of radius R, the sizes of inhomogeneous inclusions are distributed according to a power law $L^3$dn/d(ln$L$) =const, where n is the number of inclusion of size *L* in a unit volume. This distribution defines the structural self-similarity of fractals.

Let us consider that deformation starts at time $t_0$ and all stresses are zero before. Then stresses will rise at different rates for different inclusion sizes. At time *t*, there is some inclusion with size $L_M$, which reaches its highest possible stress balancing deformation and dissipation. At all bigger inclusions, stress is still growing. When the rate of deformation is high enough and there are big enough inclusions the attained stress may exceed at some point the critical stress of fracturing. Then a quake may occur. This is a transition to a super-critical regime and the sizes of earthquakes are distributed by a power law.

In economics, higher incomes are characterized by a similar distribution, but they are the net result of all forces and agents in the economy, which both vary with time. They do not represent a predefined structure as in geomechanics. Moreover, low and middle incomes are distributed according to an exponential law rather than a power one. So, we had to construct the basic distributions of defining parameters, which result in exponential distribution of low-middle incomes and power law distribution above the Pareto threshold. The process of model development with explicit differential equations together with the selection of underlying distributions is described in the following Subsections.

1.2. **Ordinary differential equation of personal income growth**

On the whole, two main driving forces of our model are similar to those in the Cobb-Douglas production function: $Y=W^\mathbf{a}K^\mathbf{b}$, where *Y* is the measure of production (*e.g.*, Gross Domestic Product) in a given country, which may be measured in the country-specific currency, *W* is the labour often considered as work hours, *K* is the physical (or work) capital (*e.g.*, machinery, equipment, buildings, hardware, software, *etc*.), and **a** and **b** are the output elasticities. Indeed, labour is the only source of products and services measured in money, and thus, the only source of income. At the same time, using larger and more efficient work instruments people produce more goods and services, also in terms of their real value measured in monetary units. This consideration is fully applicable at the level of individual production. All persons of working age are characterized by nonzero and varying among individuals capabilities to generate income and use work instruments of different sizes to do that.

Unfortunately for economics, the Cobb-Douglas function is a non-physical one. It implies the unlimited growth in GDP because it does not include any forces counteracting production. Following the physical approach discussed in Section 1.1, we assume that no one is isolated from the surrounding world. When a person starts her work some forces arise simultaneously to counteract any production action. In this setting, the work (money) she produces must dissipate (devaluate) through the entire diversity of interactions with the outside world, thereby decreasing the final income per unit time. All counteractions with outer agents, which might be people or some externalities, determine the final price of the goods and services the person produces.

Following the shape of mean income curve in Figure 1, the evolution of personal income has to be described by a phase of quasi-linear growth in the initial stage of work experience, by an exponential saturation function during the prime working age, and an exponential decline following the peak income. Given the differences between individuals, these three stages may develop at different rates. In Section 1.1, we have discusses similar trajectories and found that a larger body undergoes faster heating because it loses relatively less energy and also reaches a higher equilibrium temperature.

To characterize the change in individual income we introduce a new variable - income rate, $M(t)$, the total income person earns per year. For the sake of brevity we further call $M(t)$ "income". In essence, $M(t)$ is an equivalent of $Y$ in the Cobb-Douglas production function. The principal driving force of income growth is the personal capability to earn money, $\sigma(t)$, which is an equivalent of labour, $W$, in the Cobb-Douglas function. The meaning of the capability to earn money differs from that usually implied by the notation "human capital". Obviously, the level of human capital of many distinguished scientists is extremely high while their capability to earn money might be extremely low. Universities are full of such people. At the same time, some skills matching expectations of large audience are extremely well paid without long-term and intensive training.

Applying our physical intuition to income, we assume that the rate of dissipation of income has to be proportional to the attained level of $M(t)$. The equation defining the change in $M(t)$ should include a term, which is inversely proportional to the size of means or instruments used to earn money, which we define by variable $\Lambda(t)$. Then the dissipation term is proportional to $M(t)/\Lambda(t)$. Following the analogy in Section 1.1, one can write an ordinary differential equation for the dynamics of income depending on the work experience, $t$:

$$dM(t)/dt = \sigma(t) - \alpha M(t)/\Lambda(t) \qquad (4)$$

where $M(t)$ is the rate of money income denominated in dollars per year [$\$/y$], $t$ is the work experience expressed in years [$y$] – we have limited the maximum possible work experience in the model to 60 years, which is equivalent to 75 years of age; $\sigma(t)$ is the capability to earn money, which is a permanent feature of an individual [$\$/y^2$]; $\Lambda(t)$ is the size of the earning means, which is a permanent income source of an individual [$\$/y$]; and $\alpha$ is the dissipation factor [$\$/y^2$].

We assume that $\sigma(t)$ and $\Lambda(t)$ are mutually independent - that is a person's capability to earn money is not related to her work instrument. Notice that we have chosen $t$ to denote the work experience rather than the person's age. It is natural to assume that all people start with a zero income, $M(0)=0$, which is the initial condition for (4). At the initial point, $t = 0$, when the person reaches the working age (15 years old in the USA) her income is zero and then changes according to (4) as $t>0$. Note that both $\sigma(t)$ and $\Lambda(t)$ can vary with $t$. This means that (4) has to be solved numerically, which is the approach we apply to calibrate the model to data. Before proceeding to the calibration stage, we first make a few simplifying assumptions, under which the model has a closed-form solution.

For the sake of simplicity we introduce a modified capability to earn money:

$$\Sigma(t) = \sigma(t)/\alpha \tag{5}$$

From this point onwards we will omit the word "modified" and refer to $\Sigma(t)$ simply as to earning capability or ability. For the completeness of the model, we introduce second time flow, $\tau$, which represents calendar years. The time flow for work experience, $t$, and calendar years, $\tau$, relate to each other in a natural fashion. For a simple illustration, consider a person that turns 15 in a year $\tau_0$, *i.e.* her work experience is $t_0 = 0$. By year $\tau$ this person will have $t = \tau - \tau_0$ years of work experience. Consequently, $\tau$ is a global parameter that applies to everyone, whereas $t$ is an individual characteristic and changes from person to person.

We allow $\Lambda$ and $\Sigma$ to also depend on $\tau$, thereby introducing differences in income capability and instrument among age cohorts. In other words, the model captures cross sectional and intertemporal variation in both parameters. In line with the Cobb-Douglas production function, we make a simplifying assumption by letting $\Lambda(\tau_0,t)$ and $\Sigma(\tau_0,t)$ to evolve as the square root of the increment in the aggregate output per capita. The capability and instrument thus evolve according to:

$$\Sigma(\tau,t) = \Sigma(\tau_0,t_0) \, [Y(\tau) / Y(\tau_0)]^{1/2} \tag{6}$$

$$\Lambda(\tau,t) = \Lambda(\tau_0,t_0) \, [Y(\tau) / Y(\tau_0)]^{1/2} \tag{7}$$

where $t=\tau-\tau_0$, $\Sigma(\tau_0,t_0)$ and $\Lambda(\tau_0,t_0)$ are the initial values of capability and instrument for a person with zero work experience in year $\tau_0$; $Y(\tau_0)$ and $Y(\tau)$ are the aggregate output per capita values in the years $\tau_0$ and $\tau$, respectively, and $dY(\tau_0,t)=Y(\tau)/Y(\tau_0)=Y(\tau_0+t)/Y(\tau_0)$ is the cumulative output growth. Note that the initial values $\Sigma(\tau_0,t_0)$ and $\Lambda(\tau_0,t_0)$ depend only on the year when the person turns 15, $\tau_0$, since the initial work experience is fixed at $t=0$ for all individuals irrespective of when they start working. Now we can restrict our attention to the initial values of the capability and instrument as functions of the initial year: $\Lambda(\tau_0)$ and $\Sigma(\tau_0)$, respectively. The product of equations (6) and (7), $\Sigma(\tau_0,t_0)\Lambda(\tau_0,t_0)$, evolves with time in line with growth of real GDP per capita as in the Cobb-Douglas production function. We call $\Sigma\Lambda$ the capacity to earn money, which means that $\Lambda(\tau_0,t_0)\Sigma(\tau_0,t_0)$ is the initial capacity.

Equation (4) can be re-written to account for the dependence on the initial year, $\tau_0$:

$$dM(\tau_0,t) / dt = \alpha\{\Sigma(\tau_0,t) - M(\tau_0,t) / \Lambda(\tau_0,t)\} \tag{8}$$

Note that when we fix $\tau_0$ and restrict our attention to a person with work experience $t$, we return to our original equation (4). Moreover, the path of income dynamics depends on $\tau_0$ only through the influence of the latter on the initial earning capability and instrument; $\tau_0$ only determines the starting position of the income rate and not the trajectory of the income path, which is completely described by equation (4).

### 1.3. **Distribution of capability and instrument size**

Actual personal incomes in any economy have lower and upper limits. It is natural to assume that the capability to earn money, $\Sigma(\tau_0,t)$, and the size of earning means, $\Lambda(\tau_0,t)$, are also bounded above and below. Then they have positive minimum values among all persons, $k = 1, \ldots, N$, with the same work experience $t$ in a given year $\tau_0$: $\min\Sigma_k(\tau_0,t)=\Sigma_{min}(\tau_0,t)$ and $\min\Lambda_k(\tau_0,t)=\Lambda_{min}(\tau_0,t)$, respectively, where $\Sigma_k(\tau_0,t)$ and $\Lambda_k(\tau_0,t)$ are the parameters corresponding to a given individual. We can formally introduce the relative and dimensionless values of the defining variables in the following way:

$$S_k(\tau_0,t) = \Sigma_k(\tau_0,t) / \Sigma_{min}(\tau_0,t) \tag{9}$$

and

$$L_k(\tau_0,t) = \Lambda_k(\tau_0, t) / \Lambda_{min}(\tau_0,t) \tag{10}$$

where $S_k(\tau_0,t)$ and $L_k(\tau_0,t)$ are the dimensionless capability and size of work instrument, respectively, for the person $k$, which are measured in units of their minimum values. So far, all $N$ persons in the economy are different and at this stage of model development we need to introduce proper distributions of $S_k(\tau,t)$ and $L_k(\tau,t)$ over population.

The complete description of the development of discrete uniform distributions for $S_k$ and $L_k$ by matching predicted and observed distributions of personal income in the U.S. is presented in [Kitov, 2005b]. Here, we use the final outcome. Specifically, the relative initial values of $S_k(\tau_0,t_0)$ and $L_k(\tau_0,t_0)$, for any $\tau_0$ and $t_0$, have only discrete values from a sequence of integer numbers ranging from 2 to 30. For any person $k$, there are 29 different values of $S_i(\tau_0,t_0)$ and $L_j(\tau_0,t_0)$: $S_1(\tau_0,t_0)=2, \ldots, S_{29}(\tau_0,t_0)=30$, and similarly for $L_j(\tau_0,t_0)$, where $j=1,...,29$. Assuming uniform distribution between 29 different capabilities, we get that the entire working age population is divided into 29 equal groups. All $k$ work instruments are uniformly distributed over 29 different sizes from 2 to 30.

The largest possible relative value $S_{max}=S_{29}=L_{max}=L_{29}=30$ is only 15 times larger than the smallest $S_{min}=S_1=L_{min}=L_1=2$. In the model, the minimum values $\Sigma_{min}$ and $\Lambda_{min}$ are found to be two times smaller than the smallest possible values of $L_1$ and $S_1$, respectively. Because the absolute values of variables $\Sigma_i$, $\Lambda_j$, $\Sigma_{min}$ and $\Lambda_{min}$ evolve with time according to the same law described in (6) and (7), the relative and dimensionless variables $S_i(\tau,t)$ and $L_j(\tau,t)$, $i, j = 1, \ldots, 29$, do not change with time thereby retaining the discrete distribution of the relative values. This means that the distribution of the relative capability to earn money and the size of the earning means is fixed over calendar years and age cohorts. The rigid hierarchy of relative incomes is one of the main implications of the model and is supported empirically by the PIDs reported by the CB for the period between 1993 and 2011 [Kitov, 2005a,b; Kitov and Kitov, 2013]. The proposed uniform distributions are rather operational and should not be interpreted far beyond their usefulness to model actual distribution of personal income. For example, in this paper we lift strict assumptions of the original model in order to match the difference in income distribution between males and females. At the same time, the good fit between observations and predictions provide a solid basis to interpret observations in term of model parameters, as it adapted in physics.

The probability for a person to get an earning means of relative size $L_j$ is constant over all 29 discrete values of the size and the same is valid for $S_i$. In a given year $\tau$, all people are distributed uniformly among 29 groups of the relative ability and over 29 groups of instruments to earn money, respectively. The distribution over income involves the history of work experience $t$ described by (4), and thus, differs from the distribution over relative values. The relative capacity for a person to earn money is distributed over the working age population as the product of the independently distributed $S_i$ and $L_j$:

$$S_i(\tau,t)L_j(\tau,t) = \{2 \times 2 ,...,2 \times 30,\ 3 \times 2 ,...,3 \times 30,...,\ 30 \times 30\}$$

There are 29×29=841 different values of the normalized capacities available between 4 and 900. Some of these cases seem to be degenerate (for example, 2×30=3×20=4×15=...=30×2). However, $\Sigma$ and $\Lambda$ have different influence on income growth in (4) and each of 841 $S_iL_j$ combinations define a unique time history.

It is worth noting that our model does not predetermine actual income trajectory for real people. The model assumes that real people have incomes, which can only be chosen from 841 individual paths predefined for their year of birth. (The exception is when personal incomes reach the Pareto threshold, as discussed in the following Sections. The Pareto distribution also fixes all individual incomes, however.) This statement is equivalent to the observation that the PIDs reported by the CPS are repeated year by year, *i.e.* the portion of people in a given range of total income share is rock solid, and thus, the observed Gini ratio is constant over time.

Left panel of Figure 2 depicts the probability density function (PDF) for the distribution of the normalized capacity to earn money, $pc=S_iL_j/S_{max}L_{max}$, which ranges from 4/900 to 1. The underlying frequency distribution was obtained in 0.01 bins of personal capacity. For the lowermost incomes, we observe a local minimum. After the PDF reaches its peak value, it falls as an exponential function $0.033\exp(-2.9pc)$ between 0.08 and 0.8. In the range of the highest personal capacities, the PDF falls faster than the exponent approximating the mid-range values. In the right panel of Figure 2, we illustrate the essence of the uniform and independent distributions of *S* and *L*. We have calculated a probability density function using the PID for people between 60 and 65 years of age as reported by the CPS in 2001.

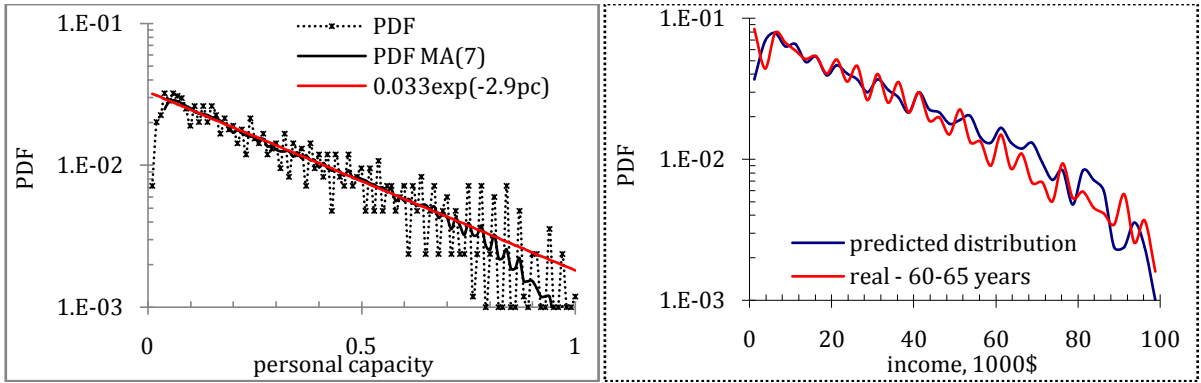

Figure 2. Left panel: The probability density function, PDF, of the normalized personal capacity, $pc=S_iL_j/S_{max}L_{max}$, distribution as defined by the independent uniform distribution of $S_i$ and $L_j$. The PDF is well approximated by an exponential function $0.033\exp(-2.9pc)$ between 0.08 and 0.8; then the PDF falls faster than the approximating exponent. Right panel: comparison of observed and predicted PDFs in 2001. The independent distribution of *S* and *L* fits the oscillations in the observed PID for people between 60 and 65 years of age.

Equation (8) suggests that many people had to reach their maximum incomes, *SL*, at the age above 60, and thus, a PDF for the real PID has to fit the theoretical distribution in the left panel. The only difference is that we have recalculated the theoretical PDF in the personal capacity bins corresponding to actual income bins of $2,500. The choice of discrete values between 2 and 30 is dictated by the fit of the observed and predicted PDFs in Figure 2. The independent distribution of *S* and *L* best fits the oscillations in the observed PID for people between 60 and 65 years of age. Any change in the range and start values (2 to 30) of $S_i$ and $L_j$ destroys the observed coherence in the PDFs' fall rate and well as the synchronization in frequency and amplitude.

Figure 3 displays the cumulative probability function, CDF, for the theoretical PDF in Figure 2. The CDF is helpful in estimation of the portion of people above any threshold. We cut top 10% of the personal capacities and found that the threshold is 0.62. For the top 1%, the threshold is 0.9. These estimates are important for further discussions of the share of people in the Pareto distribution, which is quite different from the quasi-exponential distribution below the Pareto threshold. Our model does not include any definition of "poverty" as a measure of the lowermost incomes. The CDF provides a useful tool to introduce an operational definition of a relative poverty threshold. According to the World

Bank, the relative poverty threshold is 50% of the mean income in a given country. Theoretically, the mean personal capacity to earn money is 0.283. Then the poverty threshold is 0.14. In Figure 3, red line shows that 32% of people are below the poverty line as defined by the personal capacity to earn money.

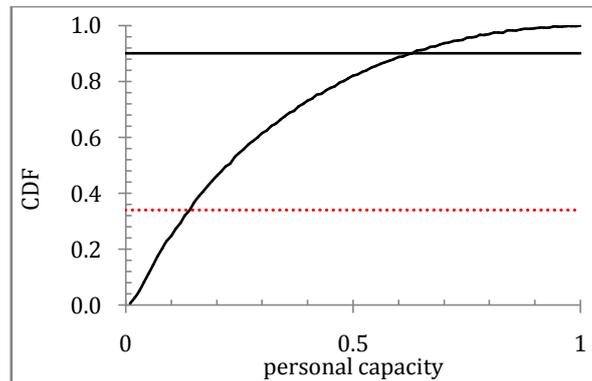

Figure 3. The cumulative density function, CDF, illustrates the rapidly decreasing portion of people with personal capabilities above some threshold: only 10% of population has the capacity to earn money above 0.62.

According the U.S. Census Bureau, the official poverty threshold in the U.S. for one person (unrelated individual) was $12,071 in 2014 and the mean income for population with income $42,789. The relative poverty threshold is then 0.08 in terms of personal capacity. It gives approximately 20% of the total working age population below the poverty line. The official level of poverty is approximately 14% of population with income. If to include 10% of population without income into the poverty statistics we obtain approximately 20% of total population as well. So, the underlying distribution of the personal capacity to earn money does predict the portion with the highest incomes and the level of poverty.

### 1.4. **Numerical modelling, personal trajectories, early rise**

Since the model contains time varying parameters, we use numerical methods to solve it and calibrate to data. However, in order to better understand the system behaviour we first consider a simplified case when $\Sigma(\tau_0,t)$ and $\Lambda(\tau_0,t)$ are constant over $t$. For shorter periods, it is a plausible assumption since these two variables evolve very slowly with time. Note that in the following exposition we fix $\tau_0$ and so income trajectories are all functions of work experience $t$ only. Given constant $\Sigma$ and $\Lambda$, as well as the initial condition $M(0)=0$, the general solution of equation (4) is as follows:

$$M(t) = \Lambda\Sigma[1 - \exp(-\alpha t/\Lambda)] \qquad (11)$$

Equation (11) indicates that personal income growth in the absence of economic growth, *i.e.* $d\Lambda/dt = d\Sigma/dt = 0$, depends on work experience, the capability to earn money, the size of the means used to earn money.

It is possible to re-arrange equation (11) in order to construct dimensionless and relative measures of income. We first substitute in the product of the relative values $S_i$ and $L_j$ and the time dependent minimum values $\Sigma_{min}$ and $\Lambda_{min}$ for $\Sigma$ and $\Lambda$. (For notational brevity we omit the dependence of parameters on time and experience.) We also normalize the equation to the maximum values $\Sigma_{max}$ and $\Lambda_{max}$ in a given calendar year, $\tau$, for a given work experience, $t$. The normalized equation for the rate of income, $M_{ij}(t)$, of a person with capability, $S_i$, and the size of earning means, $L_j$, where $i, j = \{2, \ldots, 30\}$ is as follows:

$$M_{ij}(t) / [S_{max}L_{max}] = \Sigma_{min}\Lambda_{min}(S_i/S_{max})(L_j/L_{max})\{1 - \exp(-\alpha t/[(\Lambda_{min}L_{max})(L_j/L_{max})]\} \quad (12)$$

or compactly:

$$\tilde{M}_{ij}(t) = \Sigma_{min}\Lambda_{min}\tilde{S}_i\tilde{L}_j[1 - \exp(-t(1/\Lambda_{min})(\tilde{\alpha}/\tilde{L}_j))] \quad (13)$$

where

$$\tilde{M}_{ij}(t) = M_{ij}(t) / (S_{max}L_{max})$$

$$\tilde{S}_i = S_i / S_{max}$$

$$\tilde{L}_j = L_j / L_{max}$$

$$\tilde{\alpha} = \alpha / L_{max}$$

and $S_{max}=L_{max}=30$. In this representation, $\tilde{S}_i$ and $\tilde{L}_j$ range from 2/30 to 1. The modified dimensionless dissipation factor $\tilde{\alpha}$ has the same meaning as $\alpha$ in (4).

Note that $\Sigma$ and $\Lambda$ are treated as constants during a given calendar year, but evolve according to (6) and (7) as a function of time. The term $\Sigma_{min}(\tau_0,t)\Lambda_{min}(\tau_0,t)$ then corresponds to the total (cumulative) growth of real GDP per capita from the start point of a personal work experience, $\tau_0$ ($t_0=0$), and vary for different years of birth. This term might be considered as a coefficient defined for every single year of work experience because this is a predefined exogenous variable. Relationship (13) suggests that one can measure personal income in units of minimum earning capacity, $\Sigma_{min}(\tau_0,t)\Lambda_{min}(\tau_0,t)$, for each particular starting year $\tau_0$. Then equation (13) becomes dimensionless and the coefficient changes from $\Sigma_{min}(\tau_0,t_0)\Lambda_{min}(\tau_0,t_0)=1$ in line with real GDP per capita. Further, we present simulations of individual income trajectories under the assumption of constant parameters and compare them to the calibrated version, where the output growth is taken into account and all defining parameters are allowed to grow.

For constant $L_j$ and $S_i$, one can derive from (13) the time needed to reach the absolute income level H, where H<1:

$$t_H = \Lambda_j \ln[1-H)] / \alpha \quad (14)$$

This equation is correct only for persons capable to reach H, i.e. when $L_jS_i/S_{max}L_{max}>H$. With all other terms in (14) being constant, the size of work instrument available for a person, $\Lambda_j$, defines the change in $t_H$. In the long-run, $t_H$ increases proportionally to the square root of the real GDP per capita.

Figure 4 illustrates two channels of $t_H$ dependence on $\Lambda_j$. We consider two values of $S_i=2$ and 30 and one value of $L_j=30$ and compare personal income curves in 1930 and 2011. All income trajectories span the period of 60 years of working experience. For 2011, the start year for the eldest person in the model (75 years of age) is 1951. For 1930, both trajectories begin in 1870 and follow the real GDP per capita time series. To facilitate the illustration we have divided the personal incomes by $Y(\tau)/Y(\tau_0)$ for each year before 1930 and 2011, i.e. we corrected all incomes for real economic growth.

For constant $\Lambda_j$, the time needed to reach H for a given person does not depend $S_i$ – the curves in the left and right panels are pair-wise identical in terms of shape. This means that the person with $S_i=2$ reaches, say, 50% of her maximum personal capacity $\Lambda_j\Sigma_i$ exactly at the same time as the person with $S_i=30$ reaches 50% of her maximum income. At the same time, the person with $S_i$=2 never reaches H=0.5 - her income ceiling is 1/15. As we discussed earlier, only 10% can reach the level of 0.62 and only in case they would have infinite time.

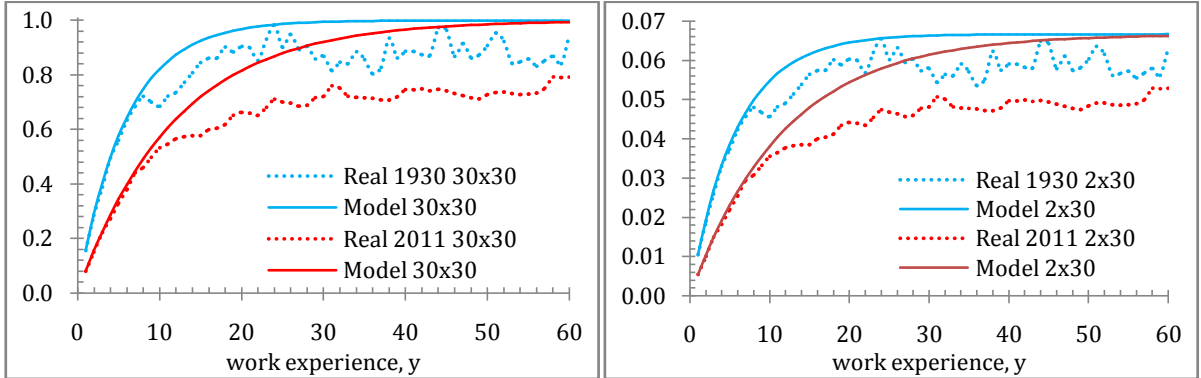

Figure 4. Individual growth trajectories for persons with two different capabilities to earn money ($S_i$) - 2 and 30, and identical $L_j$=30. The increase in $\Lambda_j$ from 1930 to 2011 results in slower income growth. Solid lines represent the solutions for constant $\Lambda$ and $\Sigma$, and dotted lines represent the numerical solution of (13) with real GDP per capita.

The increase in $\Lambda_j$ from 1930 to 2011 results in a much slower income growth. Solid lines in Figure 4 represent the solutions for constant $\Lambda$ and $\Sigma$, and dotted lines represent the numerical solution of (13) with real GDP per capita. In 1930, the person with $S=L=30$ reaches H=0.4 in 4 years of work experience and it takes 8 years in 2011. One can see that the numerically integrated curves are below the simple theoretical prediction. The increasing $\Lambda$ does affect the relative level a person can reach before the critical age discussed in Section 1.1.

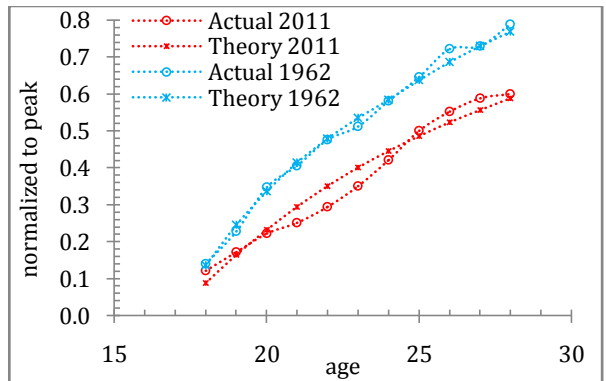

Figure 5. The evolution of normalized mean income at the initial stage. The change in growth rate with age is well predicted by the model for 1962 and 2011 as well as the change in the trajectories induced by economic growth during 50 years. At the initial stage of work experience, the input of the highest incomes is negligible – almost no people distributed by a power law. Notice that better measurements in 1962 are related to faster income growth and accompanied by a higher accuracy of prediction. In 2011, the observed fluctuations are related by poor population coverage for the youngest cohorts.

For the initial segment of income growth, when $t<<1$, the term $\alpha t/\Lambda$ in (11) is also $<< 1$. One can derive an approximate relationship for income growth by representing the exponential function as a Taylor series and retaining only two first terms. Then (11) can be re-written as:

$$M_{ij}(t) = \Sigma_i \Lambda_j \, \alpha t/\Lambda_j = \Sigma_i \alpha t \tag{15}$$

*i.e.* the money income, *M*, for a given person is a linear function of time since $\Sigma_i$ and **α** are both constants. Using (5), one regains the original meaning of the personal capability to earn money: $M_{ij}(t)=\sigma_i t$. Figure 5 illustrates the existence of a linear segment in 1962 and 2011 as well as the increase of its duration as a result of decreasing **α**/Λ, as the size of working instrument Λ grows proportionally to the square root of GDP per capita.

### 1.5. The critical age

The exponential growth trajectory of income described by equation (4) does not present the full picture of income evolution with age. As numerous empirical observations show (*e.g.*, Figure 1), the average income reaches its peak at some age and then starts declining. This is seen in individual income paths, for instance presented in Mincer [1958, 1974]. In our model, the effect of exponential fall is naturally achieved by setting the money earning capability $\Sigma(t)$ to zero at some critical work experience, $t=T_c$.

The solution of (4) for $t>T_c$ then becomes:

$$\tilde{M}_{ij}(t) = \tilde{M}_{ij}(T_c) \exp[-(1/\Lambda_{min})(\tilde{\gamma}/\tilde{L}_j)(t-T_c)] \tag{16}$$

and by substituting (12) we can write the following decaying income trajectories for $t>T_c$:

$$\tilde{M}_{ij}(t) = \Sigma_{min}\Lambda_{min}\tilde{S}_i\tilde{L}_j\{1-\exp(-(1/\Lambda_{min})(\tilde{\boldsymbol{\alpha}}/\tilde{L}_j)T_c)\}\exp\{-(1/\Lambda_{min})(\tilde{\gamma}/\tilde{L}_j)(t-T_c)\} \tag{17}$$

First term in (17) is the level of income rate attained at $T_c$. Second term expresses the observed exponential decay of the income rate for work experience above $T_c$. The exponent index $\tilde{\gamma}$ represents the rate of income decay that varies over time and is different from $\tilde{\boldsymbol{\alpha}}$. It was shown in Kitov [2005a] (and also seen in Figure 1) that the exponential decay of personal income rate above $T_c$ results in approximately the same relative level at the same age, when normalized to the maximum income for this calendar year. This means that the decay exponent can be obtained according to the following relationship:

$$\tilde{\gamma} = -\ln A / (T_A - T_c) \tag{18}$$

where *A* is the constant relative level of income rate at age $T_A$. Thus, when the current age reaches $T_A$, the maximum possible income rate $\tilde{M}_{ij}$ (for $i = 29$ and $j = 29$) drops to *A*. Income rates for other values of *i* and *j* are defined by (17). For the period between 1994 and 2002, the empirical estimates of parameters in (18) are $A=0.45$ and $T_A=64$ years (see [Kitov, 2005a] for details).

The critical age in (16-17) is not constant. For example, Figure 1 demonstrates that $T_c$ has been increasing between 1962 and 2011. Therefore, its dependence on the driving force of income distribution - real GDP per capita - has to be one of central elements of our model since any model should match the long-term observations. To predict the increase in $T_c(\tau)$ we use (14): the time needed to reach some constant income level is proportional to the square root of real GDP per capita. Assuming that the peak value of the mean income is constant in relative terms, we obtain:

$$T_c(\tau) = T_c(\tau_0) [Y(\tau) / Y(\tau_0)]^{1/2} \tag{19}$$

Figure 6 illustrates the growth in critical work experience, $T_c$, since 1930. In the original model, the initial value of $T_c(1929)=19.08$ years. The curve in the left panel illustrates time dependence and is best interpolated by a straight line with a slope of 0.28 years per year as if the real GDP per capita grows as $t^2$. In 1962, the critical age was 26.7 years (41.7 years of age) and reached 39.8 years (54.8 years of age) in 2011. During the last recession, the critical age dropped from 40.3 years in 2007 to 39.3 years in 2009. In the right panel, the dependence on GDP is shown for the same period.

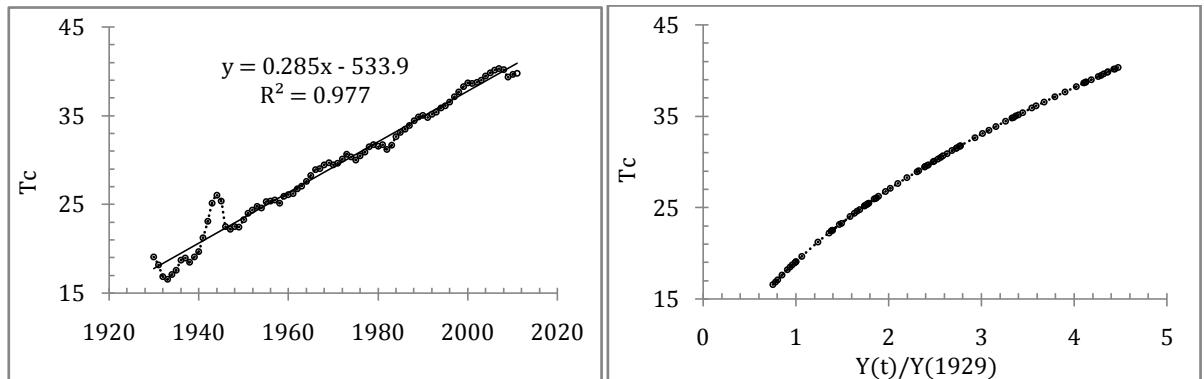

Figure 6. Left panel: Secular increase in $T_c$ is driven by the growing GDP per head. Right panel: The evolution of $T_c$ as a function of GDP growth.

Above $T_c$, people can only use their earning instrument, which is growing with time, but their capability remains at zero level and income experiences an exponential decay. Formally, the size of work instrument cannot be zero since the dissipation term would be infinite. But we can easily imagine zero capability to earn money as the absence of interest to work or work failure. The model attributes positive capability to everyone in the working age population before $T_c$. This means that each and every person in a given economy must have a nonzero income. This is not what the CPS reports – approximately 10% of the working age population reports no income from the sources included in the CPS questionnaire.

When predicting incomes we use the entire population. When comparing with observations, we include the zero-income CPS population into the income bin starting with $0 and recalculate the whole statistics like average income, the portion of people above a given threshold, etc. According to strict guidelines adapted in physics one should not calculate any aggregated characteristics of a closed system using only part of it. Such estimates are always biased and subject to fluctuations.

As an alternative to formal introduction of zero capability, one could claim that there exists a strong external process, which forces the exponential fall on top of the grown related to the original capacity to earn money. This does not resolve the problem, however, since description and explanation of these forces is needed. In addition to the homogeneous coverage of all population these forces should include the change in start time, i.e. should explain the growth in the age of peak mean income. We do not know any candidate.

Initial exponential growth and following decay, however, do not complete our model. Figure 2 shows, that our equation for income growth is not able to predict a power law distribution. We still need to introduce special treatment for the very top incomes that in multiple empirical studies have been shown to follow the Pareto distribution.

### 1.6. **The Pareto distribution of top incomes**
Our model implies that persons with the highest $S$ and $L$ may have income only by a factor of 225 larger than that received by persons with the smallest $S$ and $L$. The exponential term in (11) includes the size of earning means growing as the square root of the real GDP per capita. As a result, it takes longer and longer time for persons with the maximum relative values $S_{29}$ and $L_{29}$ to reach the maximum income rate (see Figure 4), while persons with $S_1$ and $L_1$ reach their peak income in a few years and then retain it at the level of GDP growth. The actual ratio of the highest and lowest incomes is tens of millions, if to consider the smallest reported of $1. Our microeconomic model fails to describe the highest incomes.

Fortunately, it is not necessary to quantitatively predict the distribution of the highest incomes. Here, we can adapt to our income model a concept distinguishing the below-threshold (sub-critical) and the above-threshold (super-critical) behaviour of earners. For example, using the analogy from statistical physics, Dragulesku and Yakovenko [2001] and Yakovenko [2003] associate the sub-critical interval for personal incomes with the Boltzmann-Gibbs law and the extra income in the Pareto zone with the Bose condensate. In the framework of geomechanics, adapted in this study to modelling personal income distribution [Kitov, 2005a], one can distinguish between two regimes of tectonic energy release [Rodionov *et al.*, 1982] – slow sub-critical dissipation on inhomogeneities of various sizes and fast energy release in earthquakes. The latter process is more efficient in terms of tectonic energy dissipation and the frequency distribution of earthquake sizes also obeys the Pareto law. Despite the dynamics of seismicity is not described by deterministic equations, the concept of self-organized criticality (SOC) allows reproducing statistical properties of earthquake (and not only) distribution [Lise and Paczuski, 2002].

Physics helps us to formulate an approach, which is based on transition between two different states of one system through the point of bifurcation. The dynamics of the system before (sub-critical state) and that beyond the bifurcation point (super-critical state) are described by quite different equations. It would be inappropriate to expect the equation of income growth in the sub-critical ("laminar") regime to describe the distribution of incomes in the super-critical ("turbulent") regime. In the super-critical regime, the frequency distribution of sizes (*e.g.*, magnitudes of earthquakes) is often described by a power law. It is favorable situation for our approach based on physical understanding of economy that the sub-critical dynamics can exactly predict the portion of system in critical state near the bifurcation point and the time of transition. For personal incomes, the point of transition is equivalent to some threshold, which separates sub- and super-critical regimes of income distribution.

In order to account for top incomes, which are distributed according to a power law, we assume that there exists some critical level of income that separates two income regimes: the quasi-exponential (sub-critical) and the Pareto one (super-critical). We call this level "the Pareto threshold", $M^P(\tau)$. Below this threshold, in the sub-critical income zone, personal income distribution (PID) is accurately predicted by KKM for the evolution of individual incomes. Above the Pareto threshold, in the super-critical income zone, the observed PID is best approximated by a power law. Any person reaching the Pareto threshold can obtain any income in the distribution with a rapidly decreasing probability governed by a power law. To completely define the Pareto distribution, the model for the sub-critical zone has to predict the number (or portion) of people above the Pareto threshold, which must be in the range described by the model. The predictive power of a model is determined by the possibility to accurately describe the dependence of the portion of people above $M^P$ on age as well as the evolution of this dependence over time. If the portion of people above the Pareto threshold fits observations then the contribution of the PID in the super-critical zone to any aggregate or disaggregate measure of personal income is completely defined by the empirically estimated power law exponent.

The mechanisms driving the power law distribution and defining the threshold are not well understood not only in economics but also in physics for similar transitions. The absence of explicit description of the driving mechanisms does not prohibit using well-established empirical properties of the Pareto distribution in the U.S. – the constancy of the measured exponential index over time and the evolution of the threshold in sync with the cumulative value of real GDP per capita [Piketty and Saez, 2003; Yakovenko, 2003; Kitov, 2005b, 2006]. Therefore, we include the Pareto distribution with empirically determined parameters in KKM for the description of the PID above the Pareto threshold. The stability and accuracy of the observed power law distribution of incomes implies that we do not need to follow each and every individual income as we did in the sub-critical income zone.

The initial dimensionless Pareto threshold is found to be $M^P(\tau_0)=0.43$ [Kitov, 2005a], which is within the range described by the model. Without loss of generality, we can define the initial real GDP per capita as 1. In this case, $M^P(\tau_0)=0.43$ for any starting year, where $Y(\tau_0)=1$. Then the Pareto threshold evolves with time proportionally to growth in real output per capita:

$$M^P(\tau) = M^P(\tau_0) [Y(\tau) / Y(\tau_0)] = M^P(\tau_0) Y(\tau) \qquad (20)$$

This retains the portion above this threshold almost constant over time as shown in Figure 7. In the KKM, the Pareto threshold does not depend on age.

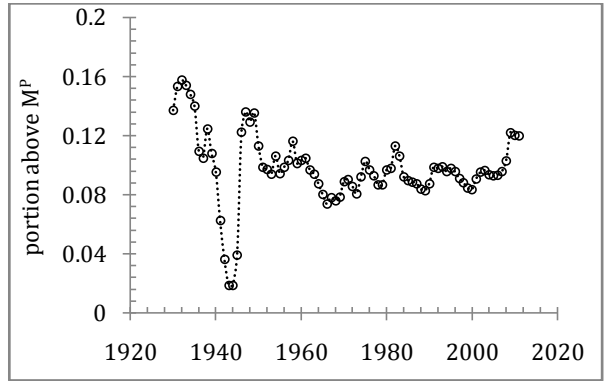

Figure 7. The portion of people above the dimensionless Pareto threshold $M^P=0.43$ between 1930 and 2011. The portion drops during WWII and hovers around 10% ever since.

As we discuss in the next Section, the Pareto threshold was different for males and females in the 1960s and 1970s. This difference is one of principal features of the gender disparity in the U.S. and deserves deeper analysis and special modelling. It is easier to incorporate the observed lower Pareto threshold for women in our model than to understand the forces behind such a difference. In this Section, we discuss the overall model and illustrate it by income features of the total population.

Theoretically, the cumulative distribution function, CDF, for the Pareto distribution is defined by the following relationship:

$$\text{CDF}(x) = 1 - (x_m/x)^k \qquad (21)$$

for all $x > x_m$, where $k$ is the Pareto index. Then, the probability density function (PDF) is defined as:

$$\text{PDF}(x) = kx_m(x_m/x)^{k-1} \qquad (22)$$

Functional dependence of the probability density function on income allows for the exact calculation of total population in any income bin, total and average income in this bin, and the contribution of the bin to the corresponding Gini ratio because the PDF defines the Lorenz curve.

The actual estimates of index $k$ reveal clear age dependence [Kitov, 2008a]. The evolution of the Pareto law index was estimated as the slope of linear regression line in the log-log scale. Using the CPS PIDs in various age groups aggregated over several years we obtained: $k=3.91$ for the age group between 25 and 34 years; $k=3.48$ between 35 and 44; $k=3.38$ between 45 and 54; $k=3.14$ in the age group between 55 and 64. It is clear that index $k$ declines with age. Obviously, a smaller index $k$ corresponds to an elevated PID density at higher incomes. The observed decrease in $k$ with age should be inherently linked to some age-dependent dynamic processes above the Pareto threshold. The declining $k$ is a specific feature of the age-dependent PIDs, which is not incorporated in the KKM yet.

For the entire population of 15 years of age and over Kitov [2008b] estimates $k=3.35$. It is close to the estimate in the age group between 45 and 54 years. This is not a coincidence since the number of people in the Pareto distribution is also a function of age and the potion of population with the highest incomes is the largest between 45 and 54 years of age. As a result, this age group has the largest input to the entire population in the Pareto range. Thus, the power law index for the entire population is practically the same as in this age group. For numerical calculations, we fix $k=3.35$ as estimated from the overall PIDs. The bias introduced by this choice into various income estimates for other age groups diminishes with their representation in the highest income range. The portion of rich people in the youngest and eldest age groups is negligible.

One can also expect that the age-dependent and overall $k$ undergo some changes over time. The overall index may vary because of the changing age pyramid and the time needed to reach the peak income $T_c$. The overall index may change because the input of various ages varies with time. Here, we study the gender-related difference in $k$, which can also be age-dependent. The observed income disparity between men and women may also be expressed in their presence among the richest share of U.S. population.

We have modelled the number of people above $M^P=0.43$ from 1930 to 2011 [Kitov and Kitov, 2013]. Left panel in Figure 8 displays the predicted and observed numbers of people above the Pareto threshold in 1962 and 2011. We have measured these numbers from the annual PIDs borrowed from the IPUMS. Here, we have to stress that we used the entire working age population as the model input and calculated the whole period between 1930 and 2014 using only real GDP per capita as defining parameters. All other constants and initial values were fixed in 1930 and their evolution was defined by GDP growth only. The microeconomic model covers more than 80 years and gives correct predictions for two randomly selected points.

The fit between the measured and predicted numbers is excellent in various aspects. First of all, two curves for 2011 are close through the entire age range, except may be the youngest ages. The theoretical curve starts from 20 years and the observed one - from 18 years of age, but the latter curve is close to zero anyway. The measured 1962 curve is slightly higher than the predicted above the peak age. Overall, the model accurately predicts the age-dependent number of people with the highest incomes in two different years. At the same time, the predictions for 1962 and 2011 are coherent in terms they are calculated in one run with the same defining parameters and exogenous parameters (GDP and age pyramid) borrowed from reliable official sources. This means that the model accurately predicts the evolution of each and every individual income and the Pareto threshold together. Taking into account the successful prediction of the past values, one may use our model for projection of income distribution in the future. The microeconomic model describes all important aspects of income dynamics.

All curves in the left panel of Figure 8 have sharp peaks and then the number of people falls to zero at the age above 75; no elder people can be found in the Pareto income zone in 1962. In order to highlight the relative dynamics above the Pareto threshold we have calculated the portion of people above $M^P$ for all ages and then normalized the obtained portion curves to their peak values. In the right panel of Figure 8 we present the normalized portion of people who has reached the Pareto threshold as a function of age. This is the best illustration of the change in $T_c$, at least the peaks are sharper than in the mean income curves (see Figure 1). The latter contain two ingredients – low-middle (sub-critical) incomes and higher (super-critical) incomes. From Figure 8, we estimate $T_c$=27 years in 1962 and $T_c$=38 years in 2011. The difference between 1962 and 2011 is 11 years. Considering the accuracy of measurements, these estimates are in a good agreement with those obtained in Section 1.5 for $T_c$ as a function of real GDP per capita. Such a big but gradual change has not been discussed in income-related economic literature yet.

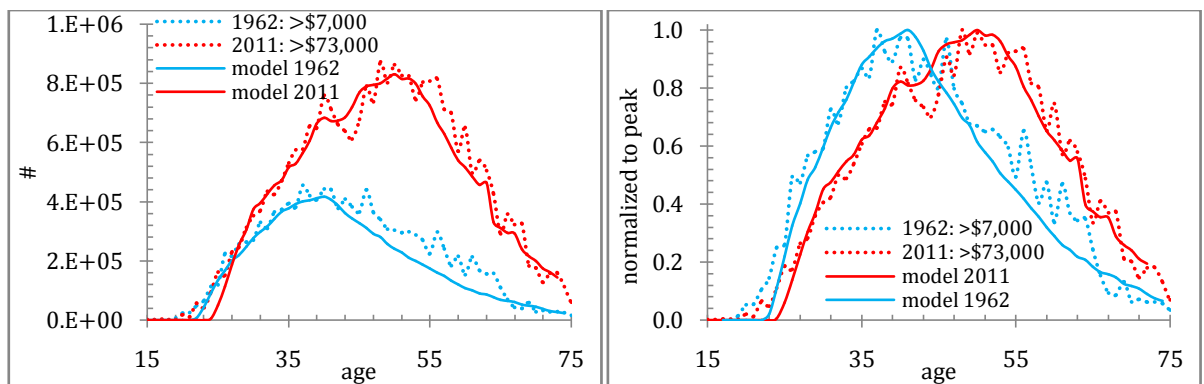

Figure 8. Left panel: The measured and predicted number of people with income above the thresholds $7,000 in 1962 and $73,000 in 2011. Right panel: The curves in the left panel are normalized to their respective peaks. The age of peak portion shifts from 42 years to 53 years.

The existence of $T_c$ proves that the personal capacity to earn money, $S_iL_j$, does define the transition through the Pareto threshold. When a personal income touches $M^P$ from below this person obtains a new (super-critical) quality to get any income from the Pareto distribution right away. However, the person retains the underlying capacity to earn money over time. Beyond the critical age, each and every personal income predicted by the microeconomic model starts to fall according to (17). At some point in time, the modelled income for a person above the Pareto threshold can break $M^P$ from above. We stress here that this is not actual income the person has by this time. It is the income she has in the model.

When a modelled income falls below the Pareto threshold, a backward transition to the sub-critical level is observed. Effectively, her top actual income momentarily drops directly below the Pareto threshold, say, from $10,000,000 to $70,000 in one year. In Figure 8, this process is expressed by rapid fall of the portion of people above the Pareto threshold beyond $T_c$: all rich people disappear above 75 to 80 years of age. The highest incomes also cannot protect a person against this transition. Otherwise, the richest persons would have higher probabilities to retain their incomes and disturb the Pareto distribution. Since our model accurately describes the dynamics of the growing and falling portion above $M^P$, we treat the personal capacity to earn money as the driving force behind the highest incomes distribution – people obtain a super-power for limited period of time and their capacities do control their incomes in the sub-critical and super-critical zones.

There is an ongoing problem with the accuracy of the highest income measurements associated with confidentiality. Since the population with top incomes is represented in the CPS universe by a few people their actual incomes are "topcoded", i.e. reduced to income bin boundaries [*e.g.*, Larrimore *et al.*, 2008]. According to IPUMS [King *et al.*, 2010], topcoding is defined as "a determination by the CPS that some high values were too sparse and specific to be recorded as they were reported to the CPS without the possibility of identifying the respondents." The bias introduced by topcoding into the mean income estimates is not the only problem. It is very unfortunate for quantitative analysis that the topcodes are prone to severe revisions by income sources and by year. Moreover, the personal income estimates above the topcode were processed in different ways over time, *i.e.* they were changed according to different rules. In any case, all these procedures result in lower incomes reported by the CPS than those in reality. The artificial difficulties related to the topcoding deserve detailed investigation [*e.g.*, Feng *et al.,* 2006; Larrimore *et al.*, 2008; Burkhauser *et al.*, 2011].

The richest people make a significant share of the total personal income and the topcoding may introduce a measurable bias in some aggregate estimates like average income. For our model, the distortion of top incomes is not relevant, however. First of all, there are age groups where the effect of topcoding is marginal. For the youngest people, the portion of people in the Pareto distribution range is negligible while the dynamics of income growth at the initial segment of work experience is the most prominent. Figure 1 demonstrates that young people raise their incomes from zero to 60% of the peak mean income in the first five to ten years. As one can see in Figure 8, the observed portion of rich population in 2011 is less than 1% for ages between 15 and 25 and then starts to grow at an elevated rate. The curves in Figure 5 evidence that the mean income of the 22-year-olds is 30% of the peak mean income measured for the 50-year-olds. As a benefit of real economic growth for quantitative modelling, the period needed to enter the top income range increases with real GDP per capita, and thus, the effect of topcoding starts at larger work experience. By good fortune, the key parameters of our model, the dissipation factor, **α,** and the minimum size of work instrument, $Λ_{\min}(τ_0)$, can be most accurately estimated using the initial segment of the growth trajectory.

Secondly, the deviation from a power law distribution and errors in income estimates related to the CPS income topcoding do not affect the portion of people above the Pareto threshold. As we discussed in the beginning of this Section, there is no physical link between these two processes in the long-term observations and in the model. Effectively, whatever process disturbs the distribution of top incomes it is not driven by and does not drive the processes of income growth/fall below the Pareto threshold. There exists only one connection between the people in the low/middle income range and the rich with the top incomes – the portion of people above the Pareto threshold. Figure 8 proves that our model exactly predicts this portion for the entire period with measurements.

The CPS observations show that the processes controlling the top incomes and those in the low-middle income are not linked. The rich and not-rich are not competing for the same personal incomes, at least for incomes from the sources included in the CPS questionnaire. The causes of the accelerated income growth in the top percentiles are in the focus of political, social as well as economic [Atkinson and Piketty, 2007; Atkinson *et al.*, 2009; Burkhauser *et al.*, 2012] discussions. They are beyond the scope of our model since the measures of income inequality are likely biased or/and misinterpreted in these discussions and they are related to the change in formal assignment of income sources to personal incomes rather than to real economic processes [Kitov, 2014].

Thirdly, the age of peak mean income, $T_c$, does not depend on the absolute value of the portion of people in the Pareto zone and the exponent, *k*, of the corresponding power law. It is defined by the sub-critical processes only. As Figure 1 proves, the peak age is the same for the CPS and IRS. This is because the sources of top incomes do not eat money from the sources of low-middle incomes.

Our model includes all necessary parameters to describe the distribution of top incomes, whatever are their sources and changes over time. We use the CPS estimates because they provide the most consistent and longer time series. As discussed above, the input of top incomes can vary with time, but such variations are fully accounted for by the changes in CPS estimates. For a quantitative model, the measured portion of true personal income [Armour *et al*., 2013, 2014] should be constant over time. The CPS data are the closest to this requirement.

### 1.7. The model: concluding remarks

Now, the model is finalized. Personal incomes in the sub-critical zone are proportional to the earning capacity $S_iL_j$ - individual income grows in time according to equation (12) until the person reaches the critical age $T_c$, above which an exponential decay according to (17) is observed. When income reaches the Pareto threshold, at any age before $T_c$, this income can take any value with the probability declining according to a power law with the empirically determined index $k$. If a given income trajectory has not reached the Pareto threshold before $T_c$, the probability to enter the super-critical zone falls to zero, because it starts to decay exponentially. Personal income above the Pareto threshold at critical work experience starts to decrease and can break the Pareto threshold from above at some point, in which case a backward transition to the sub-critical level is observed.

Every year, each single year cohort above 15 years of age is divided into 841 groups according to the capacity to earn money - that is every year the total number of 15 year-olds is divided into 841 groups, 16 year-olds are divided into 841 groups and so on. Any new generation has the same distribution of $S_i$ and $L_j$ as the previous one, but different initial values of $\Sigma_{min}$ and $\Lambda_{min}$, which evolve proportionally to the real GDP per capita. The modelled and actual PIDs both depend on the age pyramid. The population age structure is taken as an exogenous parameter. The critical work experience, $T_c$, also grows proportionally to the square root of real GDP per capita. Based on the independent measurements of the population age distribution and GDP one can model the evolution of the PID below and above the Pareto threshold.

The KKM defines the evolution of all individual incomes with all exogenous parameters borrowed from independent sources. In our previous studies, we calibrated the model to data and tested whether it can reproduce the overall PID and its aggregate features like the mean income and the portion of people above the Pareto threshold [*e.g*., Kitov and Kitov, 2013]. In this paper, we split the overall population into two approximately equal parts according to gender and retain the corresponding age pyramids and real GDP per capita as exogenous parameters driving the evolution of personal incomes. Before we calibrate two gender-related models to the March CPS income data (the IPUMS also provides CPS income microdata) we have to analyse principal properties of two PIDs and their aggregate features.

## 2. Observations
### 2.1. Personal income distribution for males and females

We first present the change in gender difference related to distribution of personal income as measured by the Census Bureau. In Figure 9, we compare the male and female population density as obtained in 1962, 1974, 1986, and 2014. The population density is the ratio of the number of people in a given income bin and the width of this bin. It is measured in the number of people per current dollar. When integrated over the entire income range, the curves in Figure 9 give the number of people. The IPUMS income microdata data are aggregated in $1000 bins between $0 and $200,000. These income bins are likely too narrow for the 2014 curves and they oscillate over the whole income range, even after smoothing with a MA(7). The same effect is observed at higher incomes, where the number of people is too small and many income bins are just empty. The scarcity of income data at higher incomes may have a negative effect on the estimates of the Pareto index.

In all years, the level of the female curve is higher at lower incomes. The female population makes approximately 52% of the total working age population, *i.e.* the number of men and women is approximately equal. Therefore, Figure 9 shows that a larger portion of women has lower incomes. The females' portion of population with the highest incomes becomes smaller and smaller with growing income. The male-female difference has been likely decreasing with time. In 2014, the males' curve is closer to the females' one than in 1962, where women had no incomes above $25,000. The male and female curves intersect at $4,000 in 1962, $7,000 in 1974, $14,000 in 1986, and at $30,000 in 2014. Beyond the crossing, the male curve deviates further and further from the female curve. The mean incomes are $9,861 for males and $4,161 for females in 1974, $21,822 and $10,741 in 1986, and in 2014 the mean income was $53,196 and $32,588, respectively.

A gender dependent feature, which will be discussed later in this Section, is the income range of the Pareto distribution. Between 1962 and 1986, the straight lines in the log-log scale reveal the Pareto range and the threshold for males was larger than for women. In 1962, the thresholds are ~$9,000 for males and ~$6,000 for females; in 1974 - above $15,000 and above $10,000; in 1986 - above $40,000 and above $30,000, respectively. This difference needs special consideration as income phenomenon as well as a problem for quantitative modelling. The lower Pareto threshold for women is a strong manifestation of disparity. The meaning of richness is different for females with income. This difference has been improving with time as two curves for 2014 demonstrate. It will take decades to fully equalize the PIDs, however.

Figure 9 highlights the effects of topcoding. In 1974, no income above $51,000 is reported and the population bin from $50,000 to $51,000 includes by a factor of 18 more males and 12 more females than the preceding bin. In 1986, the male population in the bin between $100,000 and $101,000 is by a factor of 4 larger than that in the adjacent bins. In 2015, the same effect is observed in the bin between $150,000 and $151,000, but the factor is 10. The topcoding may introduce a sizable bias in the estimate of the power law index.

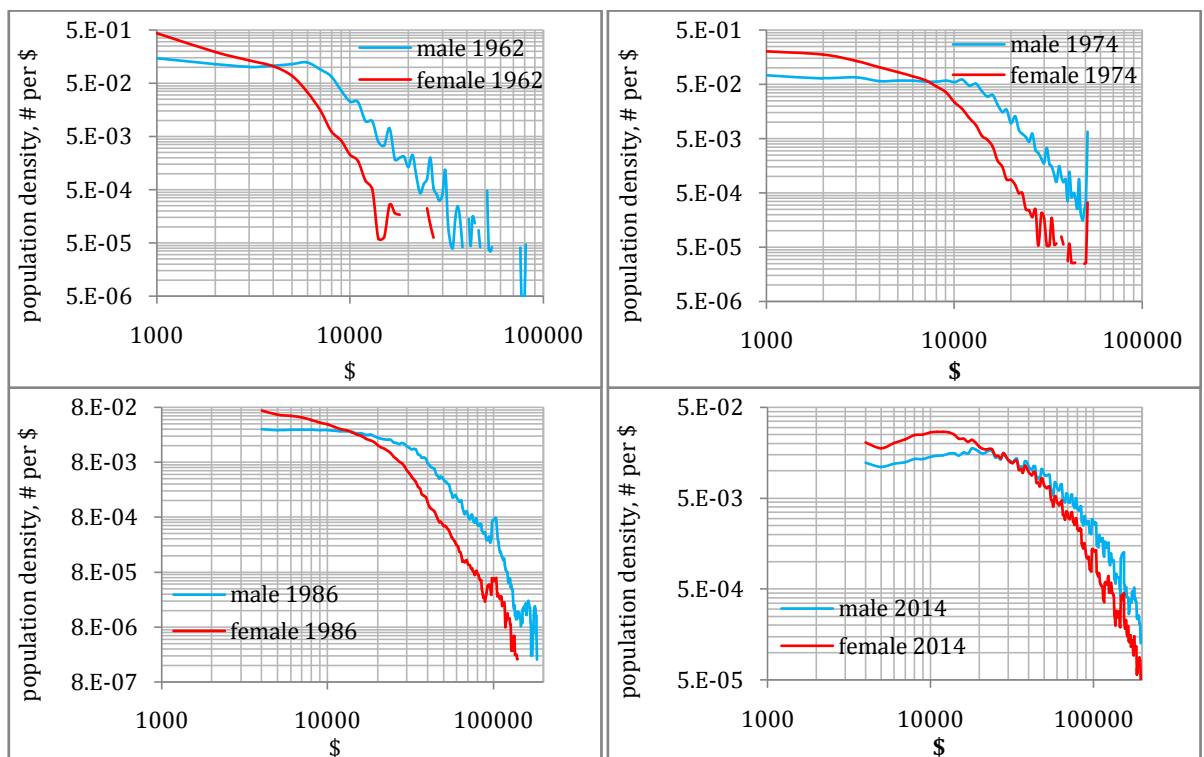

Figure 9. Population density (persons per current $) as a function of income for male and female population in 1962, 1974, 1986, and 2014. The IPUMS income data are aggregated in $1000 bins between $0 and $200,000. The 1986 and 2014 curves are smoothed by a MA(7) in order to suppress high-amplitude fluctuations. The male and female curves intersect at $4,000 in 1962, $7,000 in 1974, $14,000 in 1986, and at $29,000 in 2014.

The male and female PIDs in Figure 9 represent age-aggregated features, and thus, mask the effects of work experience. As we discussed in Section 1, the mean income and especially the evolution of the portion above the Pareto threshold both reveal strong dependence on age. The difference observed in the overall PIDs may also be associated with the change in personal income distribution as a function of age and calendar time, the latter parameter is actually the time series of real GDP per capita. The purpose of the following Subsections is to reveal the difference between genders as expressed in mean income and portion of people above the Pareto threshold and interpret them as linked to the defining parameters of our microeconomic model.

2.2. **Mean income**

The evolution of real mean income (measured in 2014 US$) for males and females is presented in Figure 10. The male curve is much higher than that for females over the period between 1967 and 2014, where the CPS historical estimates are available. The males' mean income peaks at $55,337 in 2000 and then falls to $51,119 in 2010. The females' curve peaks at $33,397 in 2007. This difference indicates that the males' and females' PIDs develop independently and react to the overall economic growth in different way.

The total curve is approximately in the middle between the gender-associated curves. It is closer to the males curve in the 1960s and 1970s, however. This can be only the result of the difference in the relative weight of males and females. Figure 11 demonstrates the portion of population with income as reported by the CPS from 1947. Here, we use the CPS electronic tables and the reports available from the CPS site as scans of the historical reports in PDF format. We have digitized all these reports and now they are available as electronic tables for quantitative modelling. The IPUMS data start in 1962.

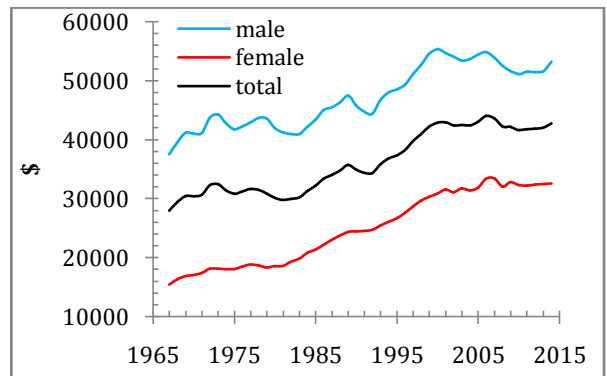

Figure 10. The evolution of mean income (real 2014 U.S. dollars) since 1967 (Source: Census Bureau, downloaded, September 25, 2015)

The portion of people with income among men does not alter much. It was 89% in 1947 and 90% in 2014. The peak portion of 97% observed in 1979 is likely a spike induced by a new methodology of income measurements. In 1977, a total revision to the CPS questionnaire as well as measuring procedure and the whole processing pipeline was implemented. In 1980, the males' portion fell back to 94% and hovered near this level before the 2000, when it started to fall to the current level near 90%.

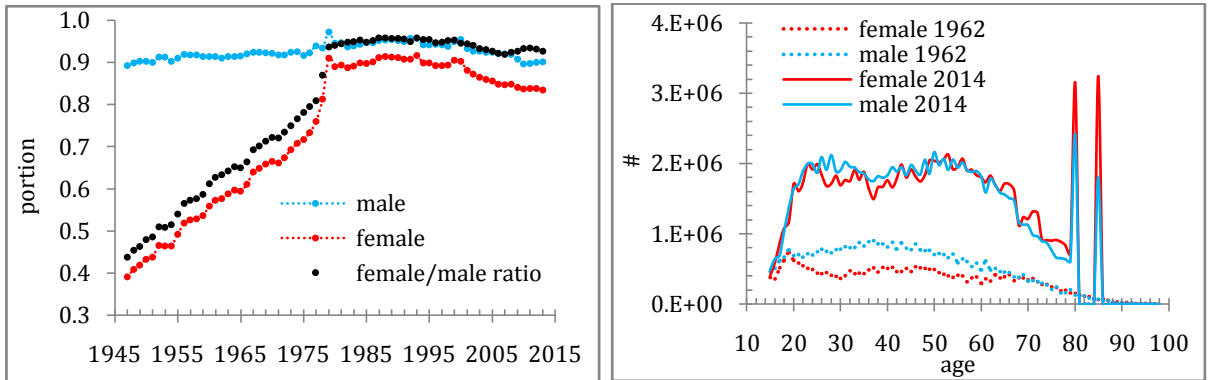

Figure 11. Left panel: The portion of male and female population with income. The female curve demonstrates a longer period of linear growth between 1947 and 1975 and then jumps by from 76% in 1977 to 91% in 1979, after a new definition of income was adopted and a new questionnaire was introduced. The black dotted line shows the ratio of female and male population with income. Right panel: The age pyramid for male and female population with income as measured in 1962 and 2014.

The females' curve starts from 39% in 1947 and grows as a linear function of time reaching 76% in 1977. As a result, the share of female population with income jumps to 91% in 1979. This is a fully artificial step, however, which should not be modelled or explained by actual mechanisms of income distribution as it is related to the measuring procedure only. As for males, the portion of females with income was hovering near 90% between 1979 and 2000, and then started to fall in 2001 to the current level of 83%. The difference between the males' and females portions is approximately 6% since 2003. The ratio of female and male portions is shown by black line in Figure 11. It was growing between 1947 and 1977 and has been around 0.93 ever since 1979.

In the right panel of Figure 11 we display the age pyramid for male and female population with income as measured in 1962 and 2014. Despite the low rate of participation of females seen in the left panel in 1962 the number of males and females is practically equal at the age of 20. A gap between genders is opening before 40 years of age and then it gradually collapses to zero at 66. In 2014, the numbers of men and women are almost the same throughout the whole period, except lower females' numbers between 30 and 50 years of age and higher ones beyond 65. The numbers of people between 80 and 85 years of age and above 85 are gathered together and create two large-amplitude peaks at 80 and 85.

Open circles in Figure 12 represent the ratio of mean incomes measured for male and female population with income since 1967. It has been decreasing from 2.43 in 1967 to 1.71 in 2014. In view of the changing portion of population with income the ratio of male and female mean income has to be corrected. Open triangles in Figure 12 present the ratio of the CPS mean incomes divided by the ratio of populations in Figure 11. In 1967, the male mean income was by a factor 3.52 larger than that measured from the female population as a whole. Our model is based on the entire working age population, and thus, the females PIDs and their derivatives measured in the 1969s and 1970s may be biased. It may be difficult to match observations made from one third or a half of population.

The evolution of age-dependent mean income curves is best represented when the curves measured in current dollars are normalized to their peak values. Figure 13 displays the normalized mean income curves for males and females for selected years with a ten-year step since 1967. The males' curves in the left panel are similar to those for the total population (see Figure 1), which is not a surprise in view of the men's dominance in the number of people with income and their higher incomes. The evolution of the males' mean income with age is also similar to that for the whole population – a short period of fast growth, which is almost linear with age, a period of saturation to the peak value at the critical age, and then all curves fall to the level between 0.3 and 0.5 at the age of 75 years. Beyond 75, the fall rate is decelerating and all curves likely collide at the level between 0.27 and 0.35. In 1997, the IPUMS income microdata are available only till the age of 90, and for the 2007 curve, microdata are limited by the age of 80.

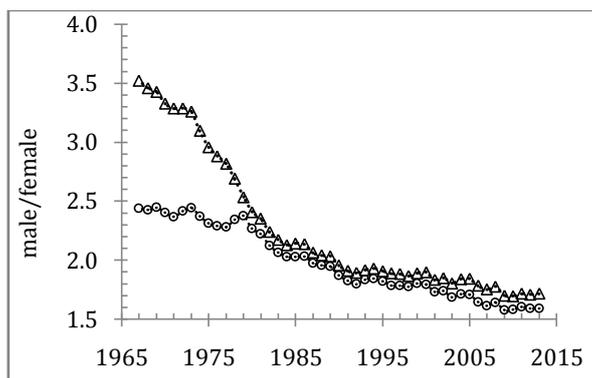

Figure 12. The ratio of male and female mean income as reported by the CPS (open circles) and that corrected to the population without income (open triangles).

The critical age measured from the smoothed males' curves in Figure 13 grows from around 40 years in 1967 to above 50 years in 2007. The change in the slope of the initial quasi-linear segment as well as the increase in the critical age is well described by the KKM. Left panel of Figure 14 compares the males' and overall mean income curves. In 1967, the overall and male curves are practically identical except a short period near 55 years of age – the influence of females is rater negligible. In 2007, the slopes of the initial segments are the same, but the peak age is slightly larger for males only. In terms of our model, males may use larger sizes of work instruments than females, and thus, than the average sizes in our original model.

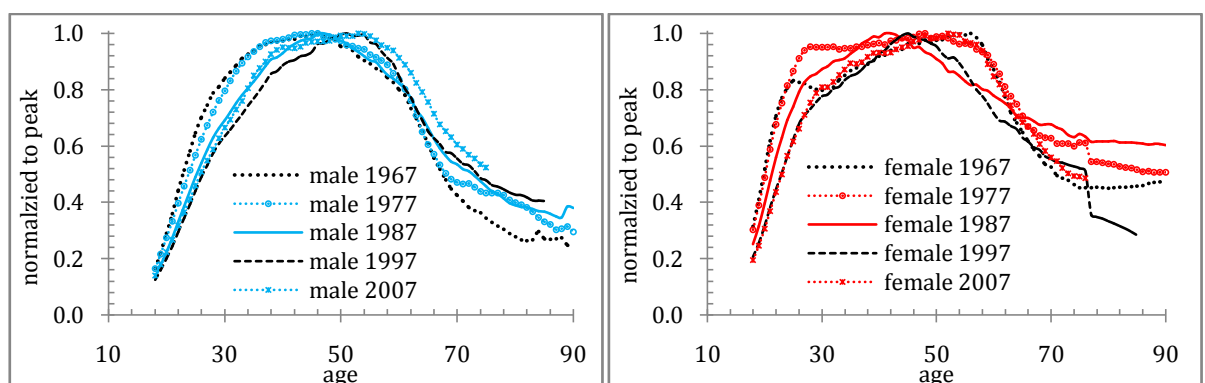

Figure 13. The mean income curves for male (left panel) and female (right panel) population. All curves are normalized to their peak values. The evolution between 1967 and 2007 is illustrated by curves with a ten-year step. For the males' and females' curves, the slope of the initial segment is falling with time and the age of the peak value grows with time. For females, the 1967 and 1977 curves demonstrates wider shelves between 25 and 60 years of age. All curves are smoothed with MA(7) before they are normalized.

The corresponding curves for females are shown in the right panel of Figure 13. Two curves for 1967 and 1977 demonstrate wider shelves between 25 and 60 years of age. This is an unusual feature not seen in the overall and male curves as reported by the Census Bureau. A wide shelf in the mean income implies no change with age, which would be hard to model considering the effect of critical age, $T_c$. A possible explanation of the shelf is the absence of people with incomes above the Pareto threshold, as discussed in Section 2.3. The evolution of incomes in the sub-critical zone suggests that they can reach their respective peaks and retain the achieved level over time before the critical age. Judging by the curves in Figure 13, the critical age for sub-critical incomes in the 1960s and 1970s is around 60 years. This value is different from the critical age measured from the portion of people with mean income as well as from the mean income for the entire population, where the input of rich males is high.

The difference between two critical ages is crucial for our model. There was no possibility to distinguish between these two critical values using only the overall PIDs. The dominance of males who have many people with incomes above the Pareto threshold may mask the difference. When modelling the females PIDs and their derivatives, we need to take into account the possibility of two critical ages. The age when the portion of rich people achieves its peak should be driven by real GDP per capita. The age when people with incomes below the Pareto threshold reach the relevant critical value can be constant.

Alternatively, it can change with the age of retirement in the U.S. The average retirement age for men has increased from 62 to 64 over the last 20 years and for women it rose from 55 in the 1960s to 62 in 2010 [Munnell, 2011]. In the right panel of Figure 13, we observe that between 1967 and 1987 incomes of elder women fall to the level between 0.4 and 0.6 at the age of 70 to 75 and then retain at this level till 90 years of age. This could be a manifestation of full retirement and permanent pension of the eldest women. For males in Figure 13, the permanent pension above, say, 75 years of age may also be the case, but the mean income is still falling because of the decreasing portion of rich people. At these ages, female population is entirely below the Pareto threshold.

In the right panel of Figure 14, we compare the overall and females curves for 1967 and 2013. The slops measured from the initial segments are quite different: the females mean income grows much faster than that of the overall population in both years, and thus, in-between. Our model implies that the initial growth is fully described by equation (13) and there exists a direct link between the slope and the absolute size of instruments used to earn money. Figure 14 suggests that the sizes of instrument available for females are smaller than those used by males. The sizes available for females were so small in the 1960s and 1970s that practically no women could get into the overall Pareto distribution, i.e. overcome $M^P$=0.43 in terms of our model.

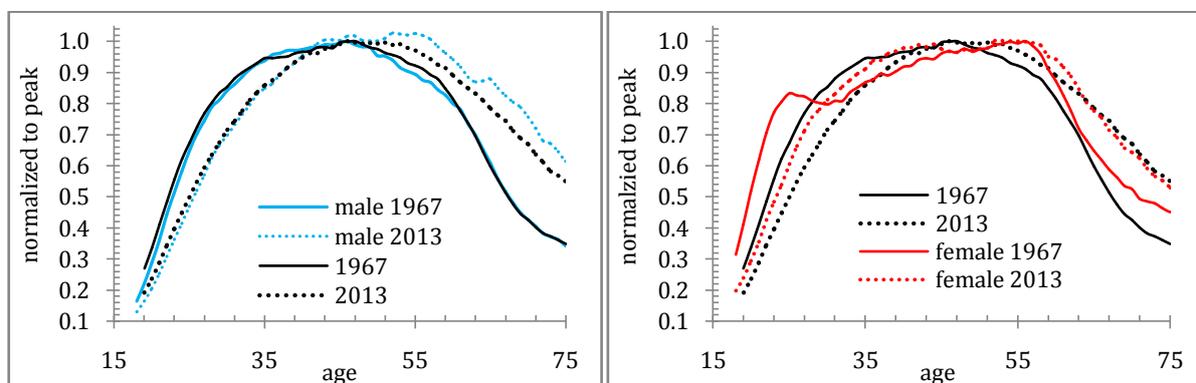

Figure 14. The evolution of mean income for male population and the overall population for 1967 and 2013. All curves are normalized to their respective maxima. The initial segments of the male and female curves are characterized by slightly lower slopes while the critical ages are shifted to larger ages.

Figure 15 illustrate the appearance of rich women near 1980 by comparison with the male curves for the same years. In 1977, the females' mean income is constant between 28 and 56 years of age. In 1982, a slight peak emerges between 40 and 45 years of age, which is closer to the critical age measured from the overall curves in Section 1. This peak is stressed by a faster fall of the females' mean income curve beyond this critical age. In 1986, the peak at 42 years of age is moderate, but it can be clearly distinguished from the males' peak around 47 years. In 1997, the males and females curves are getting more similar and two peaks now are separated by 7 years, 52 and 45 years of age, respectively. In 2007, two curves are much closer and their peak ages differ less than before.

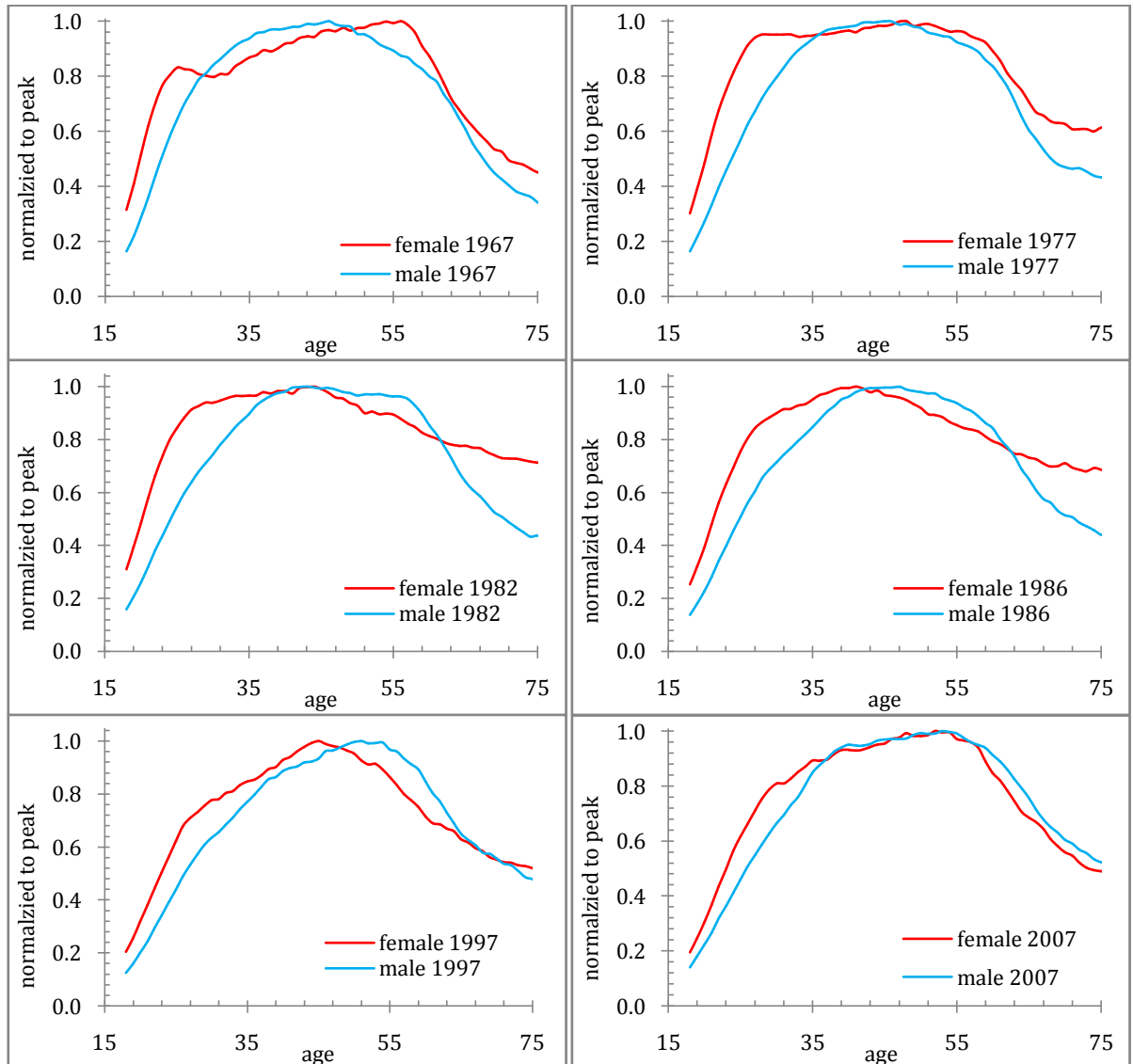

Figure 15. Comparison of the male and female mean income curves in 1967, 1977, 1987, 1997, and 2007. All curves are smoothed with MA(7)).

For all female curves in Figure 15, the initial segments of mean income growth are characterized by steeper slopes that those measured for males. This observation together with a smaller age of peak mean income indicates much smaller sizes of instruments women use to get income. The difference in the size of instruments has been likely decreasing since the start of measurements (*i.e.* 1962). Figure 16 illustrates the evolution of the male-to-female mean income ratio as obtained from the curves in Figure 13. The absolute value of this ratio decreases with time from 2.8 in 1967 to 1.87 in 2007. The estimated peak age increases from 28 in 1967 to 59 years in 2007. This transition has to be accounted for in our model, likely through changing ratio of instrument sizes available for men and women. Such a transition process was not seen when we modelled the overall data because the contribution of rich women to the mean income is small.

There is a question on the modality of instrument distribution between men and women. One possibility is that all available instruments are distributed over 29 sizes as in the original model. Men just have larger instruments from the original set and women are deprived and shifted to smaller work instruments. With time, females gradually regain their basic right to use bigger and bigger instruments from the original set. This process explains the decreasing ratio of mean incomes and convergence of the mean income curves observed in Figure 15.

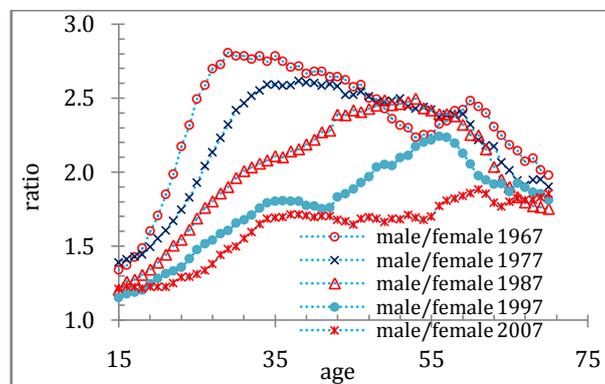

Figure 16. The ratio of male and female mean income as a function of age. The absolute value decreases with time from 2.7 in 1967 to 1.87 in 2007. The peak age increases from 28 in 1967 to 59 years in 2007.

Another option is that there are two different sets of instrument sizes – one for male and one for females. These two sets are chiefly independent and are obtained by changing distribution of the overall work capital (like in the Cobb-Douglas production function) over working age population. These should be some market forces that may differentiate males and females relative to work capital. The history of disproportionate acquisition of "human capital", whatever this term means, as well as gender prejudice are not excluded from the list of these forces. The relative distribution is changing over time and results in the observed convergence of the mean income curves.

Both opportunities are equivalent than we model male and female population separately. The sizes of work instruments are diminished for females by some factor, which has to fall with time in both cases. At the same time, modelling of the overall population as a sum of males and females described by independent models is simpler if we introduce two sets of instruments sizes. Instructively, the capability to earn money in the gender-dependent models is the same for men and women. All in all, two genders are equal in terms of their ability to earn money, but females are deprived in terms of work capital available.

2.3. **The distribution of population with income above the Pareto threshold**
As we discussed in Section 1, another aggregate variable derived from PID and sensitive to work experience is the portion of people with the highest incomes, there the highest incomes are defined as those distributed according to the Pareto law. The long-term observations in the USA reveal significant increase in the age when this portion achieves its maximum value with growing GDP per capita. Figure 17 displays five curves for selected years between 1967 and 2014. Each curve represents the age-dependent ratio of the population above predefined thresholds and total population. We have selected the following thresholds: $11,000 for 1967, $20,000 for 1977, $43,000 for 1990, $65,000 for 2001, and $87,000 for 2014. They provide approximately the same peak value of the respective ratio – 18%, which is accompanied by the same total portion of population above these thresholds - from 7.5% in 1977 to 9% in

2014. So, the integral share of population with income (male and female) above the Pareto threshold is retained at the level of 9%. One should also to take into account the deficit of female population with income before 1977. The age of peak value increases with time, as discussed in Section 1. All curves are bell shaped – for the youngest and eldest population participation in the highest incomes is negligible. The rate of growth at the initial stage is the highest in 1967 and the slowest in the 2000s. Beyond the critical age, two curves for 1967 and 1977 fall faster and reach 2% at the age of 70. The rate of fall in 2014 is the lowermost and the curve drops to 5% level at the age of 80. This observation should be reproduced by our model.

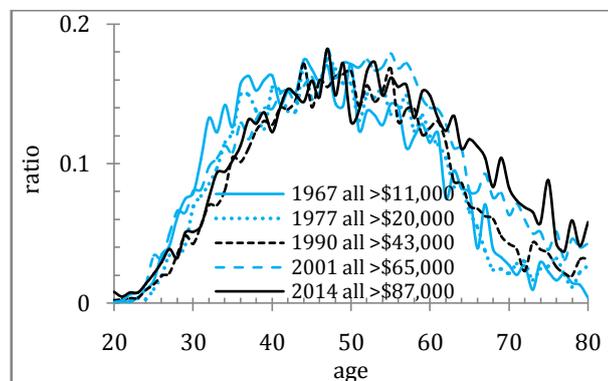

Figure 17. The portion of people above the Pareto threshold for various years between 1967 and 2014.

There are two genders in the total population, which have different contributions to the total population with the highest incomes. Figure 9 suggests that women's participation above the Pareto threshold in the 1960s and 1970s was extremely low. Figure 18 displays curves similar to those in Figure 17 with the same thresholds, but for two genders separately. In 1967, the share of rich females was less than 3% of the number of women with income, which was approximately 60% of the total number of females. Since the contribution of women was so low, male population occupied almost all positions in the high income range. Therefore, the curves in Figure 17 and the males' curves in Figure 18 do not differ for 1967 and 1977; except the males' peak portion is 28%. The women's share has been increasing since the earlier 1980s and approximately 10% of women between 35 and 65 years of age had incomes above $87,000 in 2014. These women displaced men from the top income zone and only 22% to 24% of males had incomes above the same threshold.

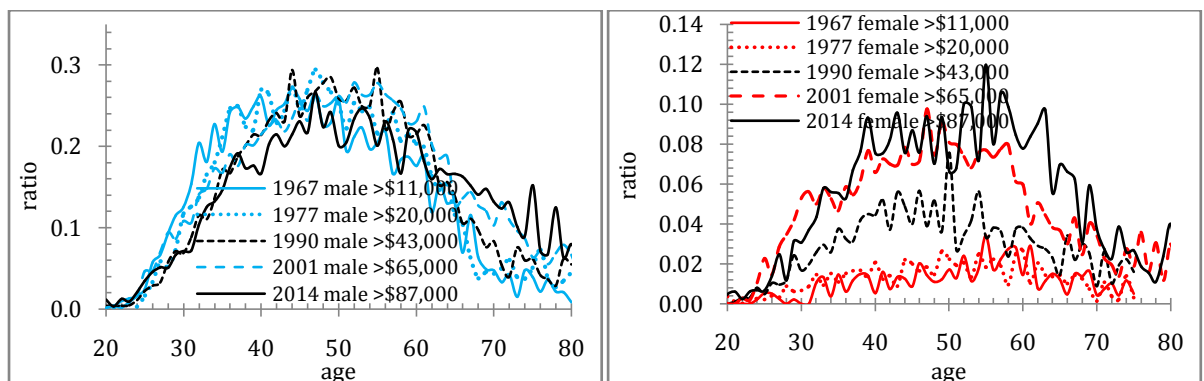

Figure 18. The portion of males (left panel) and females (right panel) above the Pareto threshold for various years between 1967 and 2014. Thresholds are the same as in Figure 17.

The females' contribution is still lower and one can estimate the time when the current convergence tendency will end in equal representation. Figure 19 shows the age-dependent ratio of the curves in the left and right panels of Figure 18. In 1967, the ratio peaks at 30 years of age and achieves the level of 30 and above. In 1980, the ratio hovers near 10 and then drops to 6 to 8 in 1990. In the 21$^{st}$ century, the ratio falls to 3.5 in 2001 and currently is between 2 and 3. This trend is a quasi-exponential one ($R^2$=0.98) and the extrapolated curve will reach 1.0 in 2025-2030. This is the expected time of gender equality as related to participation in the Pareto distribution. The trend may change in the future, however. To model the past numbers of females above the Pareto threshold we may adjust the relevant defining parameters to fit the exponential fall in the male-to-female ratio.

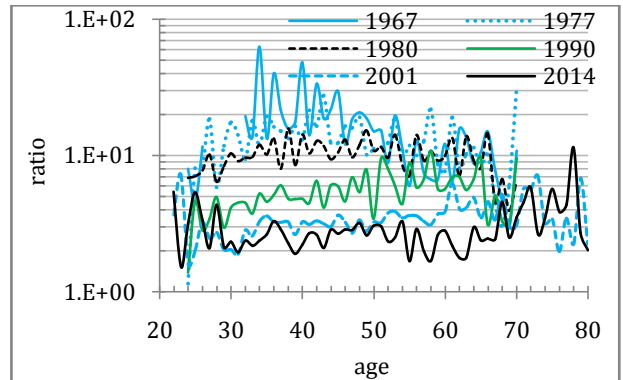

Figure 19. The ratio of male and female portions in Figure 18.

Figure 20 illustrates the evolution of age dependence with time and compares male and female curves. We have estimated the age-dependent curves for the portion of population above given thresholds similar to those in Figure 18 and normalized them to their peak values. Male and female curves are compared on a year-by-year basis. The thresholds in Figure 20 are lowered in order to obtain more reliable estimates with smaller fluctuations. The negative effect of the decreased thresholds is that they are now below the Pareto one for males but still above that for females.

The total portion of people above these thresholds varies from 13% in 1977 to 17% in 2014. In 1967, the males curve grows at a high rate from 23 years of age, peaks at 35 years of age, and then falls to 0 at 75 years of age. The female curve grows slowly with large fluctuations and has a sharp peak at 55 years of age. In 1977, the female curve peaks at 47 years of age and has a plateau till 65 years. After a period of fast growth in the females' share between 1977 and 1980, we observe that the male and female curves become closer and closer from 1990 to 2014. This is only relative convergence, however. The absolute levels differ by a factor of 2.5 in 2014. Women are as efficient as men when they get in the top income percentiles. They are underrepresented however.

The discrepancy between the male and female curves in Figure 20 observed in 1967 and 1977 may be induced by the difference in the Pareto thresholds. For females, it is much lower, as Figure 9 shows. In Figure 21 we display the age-dependent portions of females with income above $6,000 in 1967 and $14,000 in 1977, instead of $9,000 and $17,000, respectively, in Figure 20. The peak values are now around 20% of females with income; it has to be reduced according to the total share with income below 70% (see Figure 11). In terms of shape, both curves are now similar to those in Figure 15, which presents the mean income curves.

We know from Figure 9 that men and women in the U.S. have different Pareto thresholds. Moreover, these thresholds change with time at different rates. In order to better illustrate the difference between males and females, Figure 22 displays only PID segments which are characterized by power law for selected years between 1967 and 2014. The start points of these segments provide more accurate estimates of the Pareto threshold. For example, in 1967 the males' threshold is approximately $9,000 and only $6,000 for females.

In 2014, the Pareto thresholds are very close, if distinguishable at all, $79,000 and $78,000 for men and women, respectively.

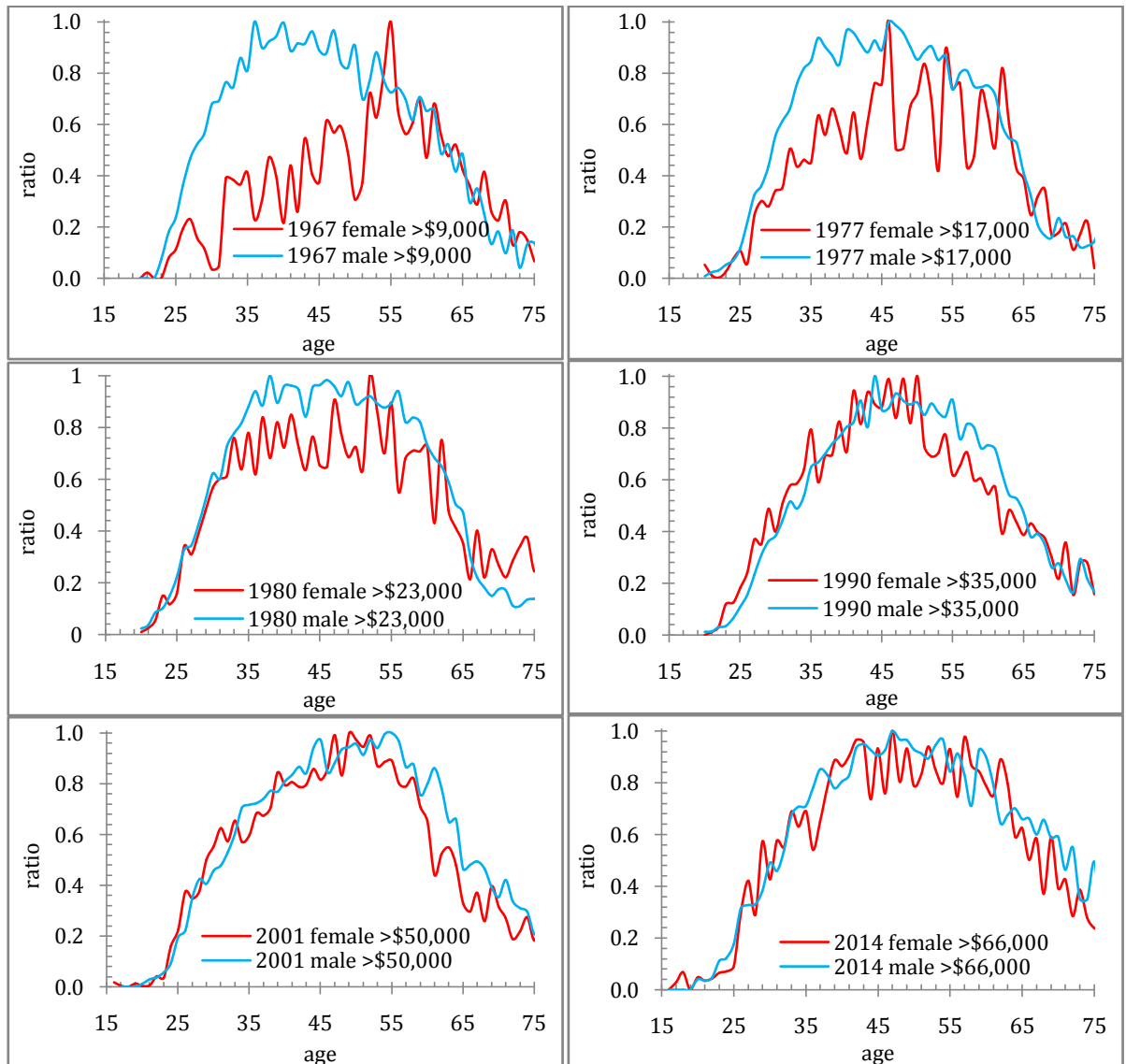

Figure 20. Pair-wise comparison of male and female curves for selected years between 1967 and 2014.

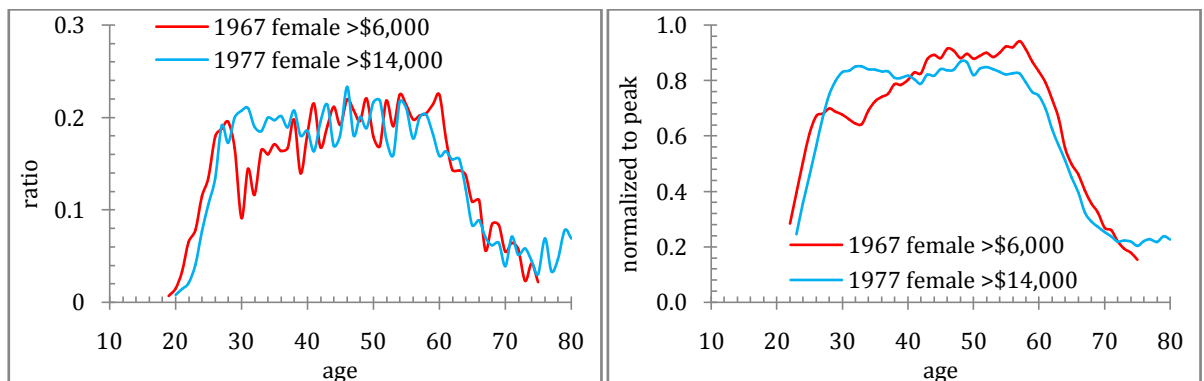

Figure 21. The age-dependent potion of females in 1967 and 1977 when the Pareto thresholds are decreased to the level corresponding to women, as illustrated in Figure 9. Left panel: absolute portion. Right panel: normalized to peak value and smoother with MA(7).

From all curves in Figure 22, we have removed the portions with top incomes because of high amplitude oscillations which spoil the power law behaviour. In addition to more accurate estimates of the thresholds themselves one can observe the difference in the power law index between male and female population. This index is estimated using standard regression technique, which we present in the log-log scale. The linear regression lines provide slope estimates, which are identical in absolute value to the power law index, *k*. All regression lines are characterized by high $R^2$ – from 0.897 for males in 2014 to 0.995 for males in 1977. The PIDs in the 21$^{st}$ century reveal clear signatures of topcoding and sparse data in the top income range.

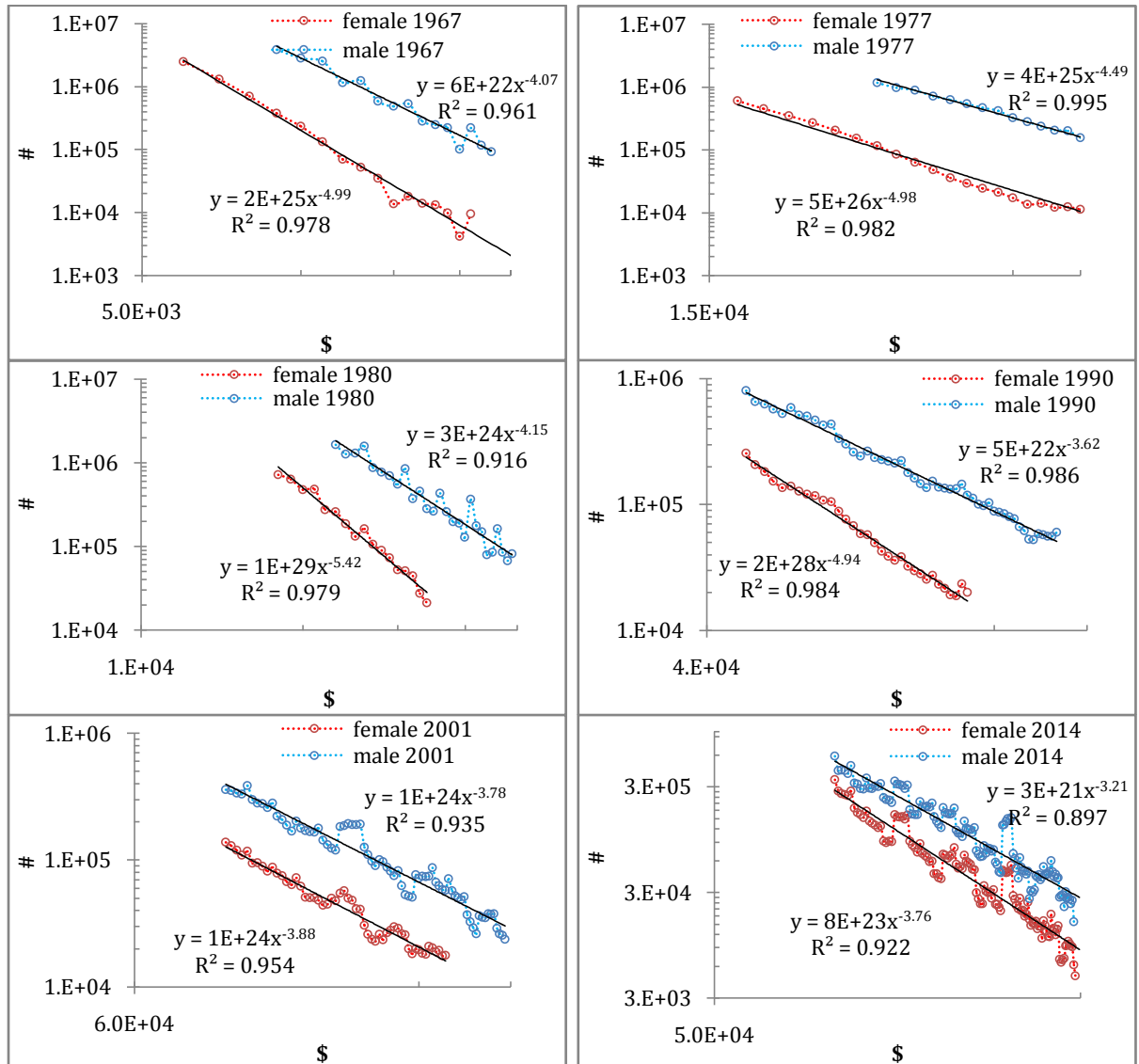

Figure 22. Personal income distributon beyond the Pareto threshold. The power law index for male and female population between 1967 and 2014 is estimated using standard linear regression in the log-log scale. The slopes of regression lines are the power law indexes.

The power law index varies in a wide range from 5.42 for females in 1980 to 3.21 for males in 2014. In the original model, we fixed the index to 3.35, which is between the values for males observed in 2001 and 2014, and obtained relatively accurate predictions of the PID, mean income, the portion of people above the Pareto threshold, and finally, the Gini ratio [Kitov and Kitov, 2013]. The discrepancy between the observed indexes and that used in our model is not harmful, however, because male population of the prime age dominates in the high-income range. The curves in Figure 22 show that the males' index is always lower than that for females in the same year. This implies that the ratio of the number of men and women having incomes above some high level increases with the level. The male/female disparity progresses with income, but it is getting lower with time.

To model male population, we do not need to change the power law index used in the original model. They are practically identical. For female population, the index estimated from the overall PID may bias the result of mean income prediction since the input of rich women would be overestimated. For calculations of the portion of people above the Pareto threshold, index $k$ is not relevant because this portion is fully defined by the sub-critical income behavior.

## 3. Modelling genders
### 3.1. The model flexibility

Our microeconomic model has a large degree of freedom despite it is driven by one exogenous variable – real GDP per capita. One can change the distribution of the capabilities to earn money and the sizes of work capital or scale them in a different way. The dissipation factor, $\alpha$, has to be estimated from data as all other constants and variables in the model, but one can also change it according to own understanding of income discounting. The critical age, $T_c$, can be changed as well as its functional dependence on GDP. The index of exponent describing the fall beyond the critical age is difficult to estimate because of data scarcity and low accuracy as observed for the youngest and eldest population. All parameters are adjustable and do change the model outcome. But this change is not random and cannot fit artificially designed personal income distribution. The changes associated with the adjustment of model parameters have to fit actual observations in the U.S. and do fit these observations.

For the overall population, we have estimated the best-fit set of constants and variables. Any change reduces the fit with observations. At the same time, the male and female PIDs, their derivatives and aggregates are so different that some changes in defining parameters of the original model are inevitable. Moreover, there are a few new features associated with the female income distribution, which we did not observe in the overall PID. These features have to be included in the model, but they should not disturb the results of the original model. In this Subsection, we demonstrate how the original model reacts to the change of some defining parameters in view of the new features discussed in Section 2.

In this study, we consider males and females separately. They are different in the age pyramid, which is an important exogenous parameter of the model. Crudely, women make approximately 51% of the total working age population in the U.S., and thus, the males' share is 49%. For the sake of simplicity we multiplied the overall population pyramid by factor 0.51 and 0.49 and obtained the female and male age pyramid, respectively. This approach may introduce some minor errors in the gender ratio for some ages, but these errors are smaller than those related to income measurements and GDP estimates. And the population pyramid does not affect the mean income predictions, which are based on the proportional distribution of people over the capability to earn money and the sizes of work instruments.

The initial value of the critical age, $T_c$, is borrowed from the overall model and is fixed to 19.07 years in 1930 in all models [Kitov and Kitov, 2013]. In some versions of the gender-dependent model we move the start year to 1960. Then the initial value of critical age is changed according to (19), i.e. as the square root of the cumulative change in the corrected real GDP per capita. It makes 25.92 years in 1960. This value is also fixed in the model. The index of the Pareto law is fixed to $k=3.35$ for both sexes and does not depend on calendar years and age. As we found in the previous Section, the largest deviations in $k$ are observed for ages having extremely low representation in top incomes. Females' participation in the Pareto zone is low throughout the entire 20[th] century. As a result, constant $k$ does not affect the accuracy of model predictions.

In Section 2, we discussed a possibility that the size of instruments used to earn money may be smaller for women than for men. The instrument size affects the time needed to reach a given threshold as well as the level of income one can obtain. In order to change the instrument size we introduce a scaling factor, FL, and multiply all standard sizes by this factor. Figure 23 presents the evolution of predicted mean income in 1962 and 2011 for three different cases: FL=0.5, 1.0, and 1.5. In order to retain the portion of population above the Pareto threshold at constant level we scale standard $M^P$=0.43 by the same factors as the instruments. As a result, $M^P$ in Figure 23 has values 0.215, 0.43, and 0.645 for FL=0.5, 1.0, and 1.5, respectively.

The increase of all standard sizes $L_j$ by factor 1.5 results in a larger relative income, which is just scaling the mean income curve: the ratio of the peak incomes in 1930 and 2014 is retained in all three cases. The critical age $T_c$ also does not change because it depends only on GDP. The slopes of two FL=1.5 curves in Figure 23 decrease relative to those in the standard model with FL=1.0, however. This is the same effect as observed with increasing real GDP per capita in the overall mean income curves (Figure 1). All in all, the increased sizes of work capital do not create new effects.

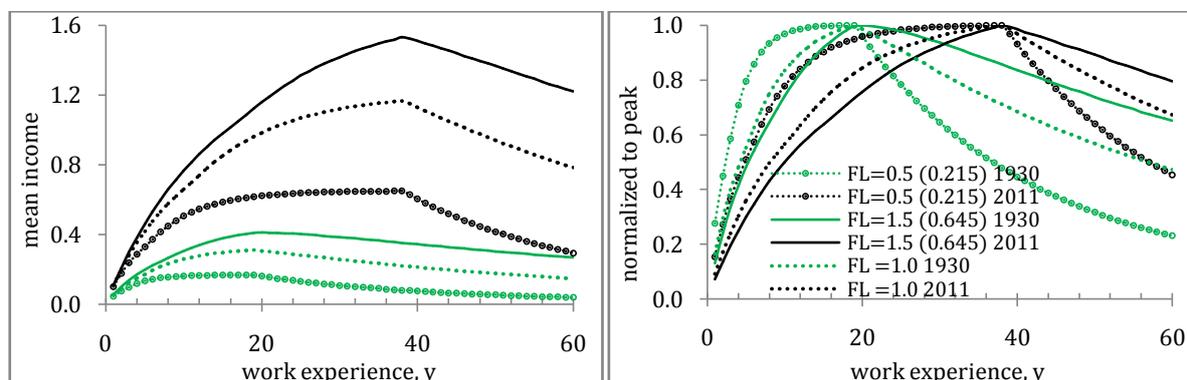

Figure 23. Mean income as a function of age in 1930 and 2011. The size of work instrument is multiplied by a factor FL=0.5, 1.0, and 1.5. The Pareto thresholds $M^P$ are also multiplied by the same factor in order to retain the portion of people with the highest incomes. They are 0.215, 0.43 and 0.645, respectively. Notice a shelf in the mean income curve for 1930 with FL=0.5.

For smaller sizes, the mean income curves become different. They contain periods when mean income does not change: between 5 and 18 years of work experience in 1930 and from 20 to 37 years in 2011. This is the effect we have found in the females mean income in the 1960s and 1970s (see Figure 15). As we know from the discussion in Section 1, smaller instruments imply faster growth of all incomes. The initial segments of mean income demonstrate steeper growth to maximum values for all earning capabilities $S_i$ from 2 to 30. The lowered Pareto threshold suggests that everyone who could reach it in normal conditions FL=1.0 does achieve it, but much faster. When all people reach their maximum incomes, including those in the Pareto zone, the model suggests no further changes before the critical age. With growing GDP, the time when all incomes reach maximum and $T_c$ both increase. The start point and duration of two shelves in Figure 23 both increase.

The portion of people above the Pareto threshold is presented in Figure 24. Both curves for FL=0.5 are characterized by early and steep growth. The number of people is displayed in the left panel. Here, we use the males' age pyramid and all observed fluctuations are associated with the varying number of population rather that with the model parameters. It is instructive that the total number of people increases with falling FL – faster income growth involves younger population into the Pareto zone. The number above $M^P$ in a given age does not affect the mean income since the portion does not depend on the total number. That is why the mean income curves in Figure 23 are smooth.

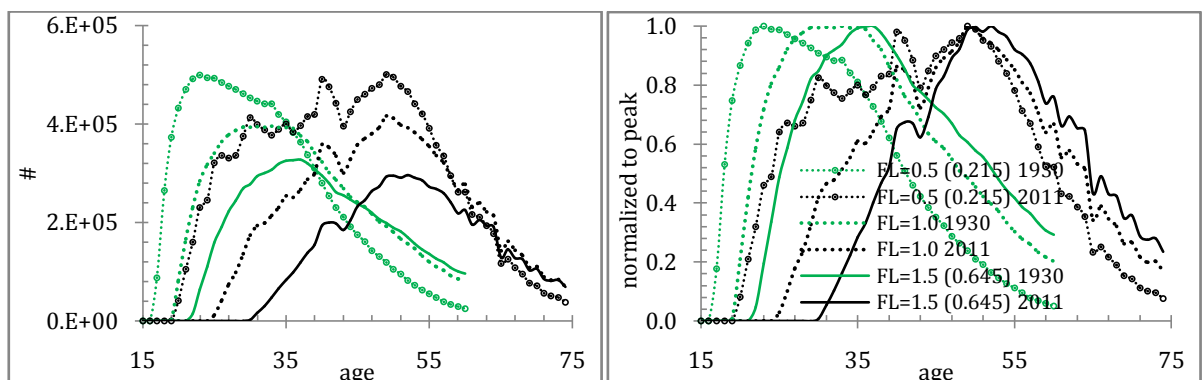

Figure 24. The number (left panel) and portion (right panel) of people above the Pareto threshold as a function of age in 1930 and 2011. The size of work instrument is multiplied by a factor FL=0.5, 1.0, and 1.5. The Pareto thresholds $M^P$ are also multiplied by the same factor in order to retain the portion of people with the highest incomes. They are 0.215, 0.43 and 0.645, respectively.

When the Pareto threshold is retained at the same level for all FL, the number of people is much lower for FL=0.5, as Figure 25 demonstrates. This might be the reason of very low women's representation in the top incomes in the 20$^{th}$ century. For FL=0.5, the number of people is just marginally above zero in 1930 and 2011, as was observed in the females' distribution in Figure 18. So, the lowered sizes of work instruments available for women in the U.S. result in a very low number of women with top incomes.

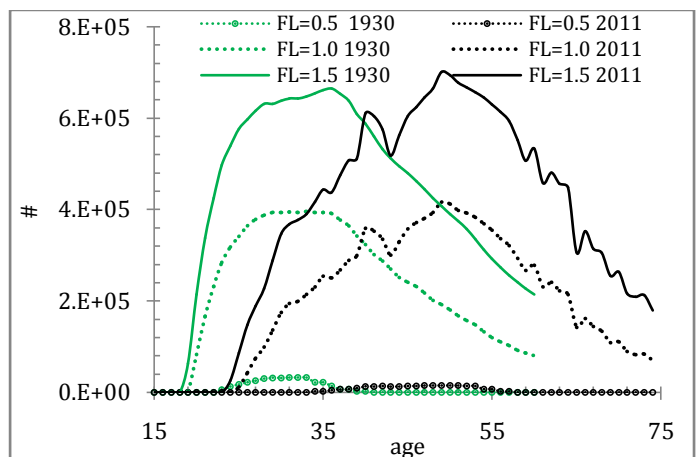

Figure 25. The number of people above the Pareto threshold as a function of age in 1930 and 2011. The size of work instrument is multiplied by a factor FL=0.5, 1.0, and 1.5. The Pareto thresholds $M^P$=0.43 for all cases.

In Section 3.2, we demonstrate that by change in the size of instrument people use to earn money and synchronized change (or no change) in the Pareto threshold our model is able to qualitatively explain some striking differences in income distribution as observed for men and women. This is a good basis for accurate quantitative prediction of principal features observed in the males' and females' PIDs and their derivatives. The age dependence of the mean income and the portion above the Pareto threshold are likely the most prominent features which demonstrate secular evolution coherent with the growth in real GDP per capita. At first glance, male population in the U.S. demonstrates simpler behaviour. We begin gender-dependent modelling with prediction of men's personal incomes.

### 3.2. **Male model**

There are some new features we can predict using the original setting of our model. Figure 14 shows that the male mean income curves have the same critical age and rate of growth as the total population. In 2013, the critical age is slightly larger for men. One may suggest that men in the U.S. economy use larger work capitals than women. In essence, their work instruments create the size distribution defined in Section 1. Within our framework, men's incomes can be modelled with standard instruments.

In Figure 26, we present the predicted and observed mean income curves for 1962, 1977, 1987, and 2011. All defining parameters are the same as in the original model together with 1930 as the start year. There are no income microdata before 1962 and we compare our predictions with actual measurements. The fit between the curves depends on age and year. For the model, the most important part of income evolution is before the critical age. This segment is the most sensitive to defining parameters including the critical age. Therefore, the almost perfect match observed before the critical age during the period from 1962 to 2014 proves that our model predicts personal incomes of male population in the U.S. with incredible accuracy.

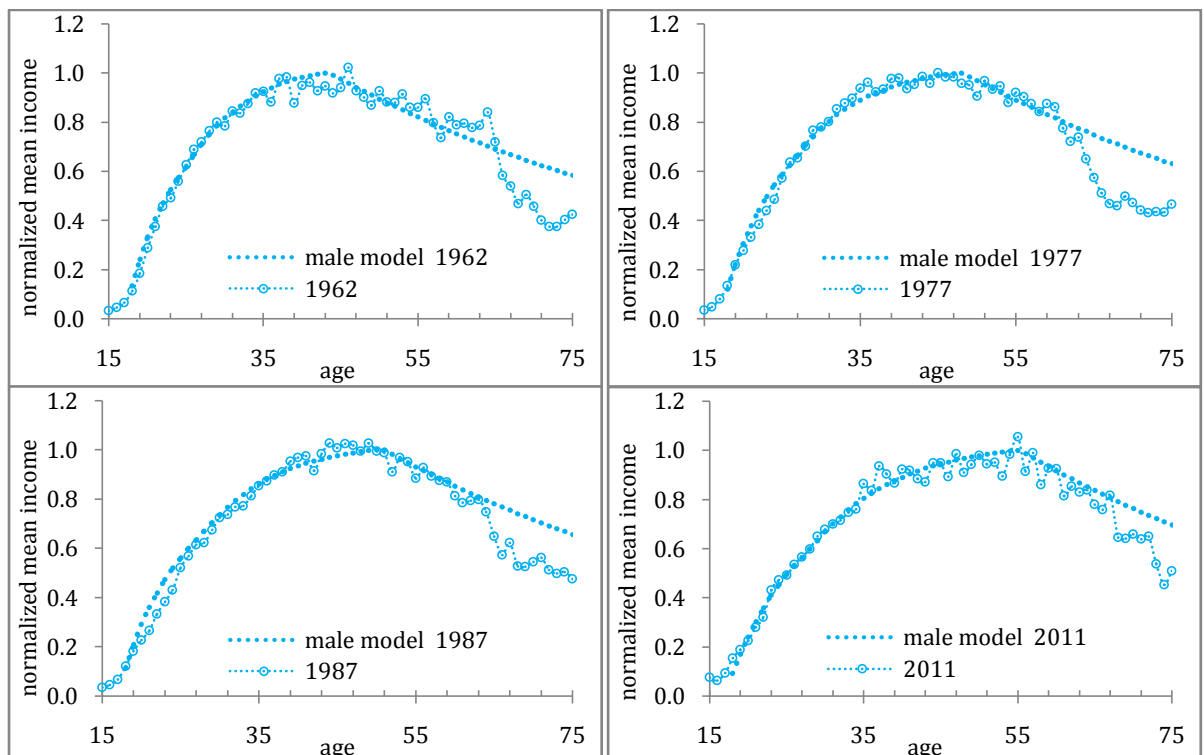

Figure 26. Comparison of measured and predicted mean income as a function of age. Selected years between 1962 and 2011 are presented. The measured and predicted curves start to diverge above 64 in 1962 and above 68 years of age in 2011.

The change in the slope and shape of the initial segments is described precisely as a function of GDP. Real economic growth leads to slower relative income growth as was already found in the overall model. The critical age, *i.e.* the age of peak mean income, has been growing since 43 years of age in 1962 and currently is above 56 years of age. We expect further increase in the age of peak income.

There is a common feature observed in all measured curves above the critical age, which is best highlighted when compared to the predicted curves following the exponential fall defined by $T_A$=60 years of work experience and A=0.65 (see equation (18) for details). At the age from 64 in 1962 to 68 in 2011, the measured curves experience a sharp drop to the level between 0.4 and 0.5. The period between $T_c(t)$ and this specific age decreases with time – from 20 years in 1962 to 12 years in 2011. In Figure 13, this effect was ironed out by smoothing with MA(7). As we discussed for female population, the reason behind this drop is likely associated with retirement. The retirees have constant income as a share of their work incomes. Then the time and amplitude of the drop can be explained by the performance of social security system.

The overall fit between the predicted and observed mean income suggests that the portion of people above the Pareto threshold should also be accurately estimated. The overall model exactly predicts the number of people of a given age above the Pareto threshold. One can directly calculate their total sub-critical income as a sum of all $M_{ij}(\tau,t)>M^P$, as if they did not move to the super-critical regime of income distribution. The net gain obtained by these people when they move to the super-critical power law distribution can be calculated since we have the number of people for each age and the overall index $k$.

Obviously, the net gain is constant for a given $M^P$ since the sub- and super-critical total incomes are fixed. For the original model, the ratio of the super- and sub-critical total incomes is 1.33 for $M^P$=0.43 [Kitov, 2009]. As a result, we do not need to calculate individual incomes in the super-critical power law distribution to get an estimate of the total contribution of rich people to the mean income. We just need to multiply all $M_{ij}(\tau,t)>M^P$ by a factor of 1.33. Hence, when the mean income dependence on age is accurately predicted we believe that the number of people above the Pareto threshold is also accurately described as a function of age.

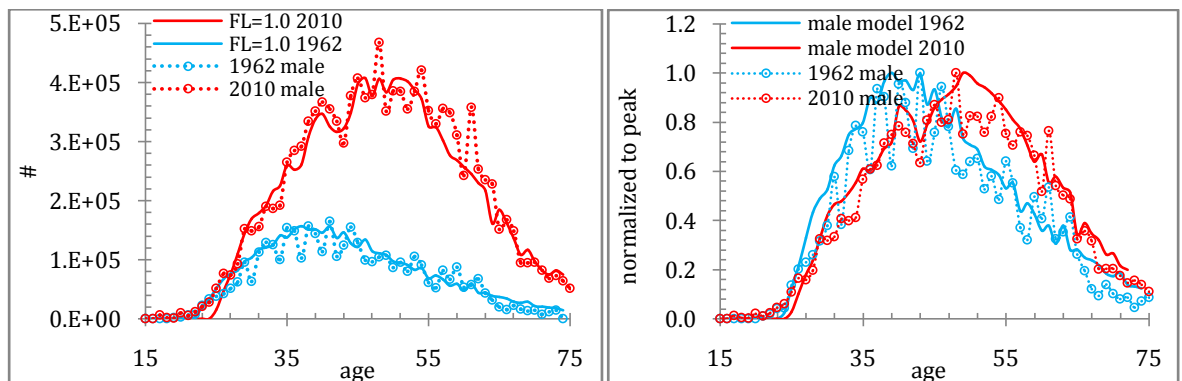

Figure 27. Comparison of measured and predicted number of people above the Pareto threshold. Actual thresholds are $11,000 in 1962 and $87,000 in 2010.

Figure 27 illustrates this statement. In the left panel, the measured and predicted number of males is presented as a function of age for 1962 and 2010. The overall fit is more than excellent considering only one parameter (real GDP) describing the whole variety of changes (*e.g.*, inflationary periods, recessions, high and low oil prices, changes in fiscal and budgetary rules, varying accuracy of all involved measurements, revisions to all involved variables among many others) during the past half-century. Moreover, to predict the number of the elder males the model involves their almost 60 year history of work experience, which includes the real GDP per capita time series since 1900. Figure 4 illustrates the effect of work experience history on income in a given year. In the right panel of Figure 27, we demonstrate the difference in the critical ages. Unfortunately, the age estimates are subject to bias because of high-amplitude fluctuations, which are related to the data scarcity at higher incomes and, partially, are induced by topcoding.

The drop at the age of retirement is not seen in Figure 27. This observation suggests that retirement does not affect the processes in the Pareto distribution. At the same time, the fall in the number of people above the threshold beyond the retirement age is accurately predicted by our model, which uses the capability to earn money and the size of work capital as defining parameters evolving as the square root of the real GDP per capita. Therefore, the effect of retirement is likely an artificial feature related to sub-critical incomes, which should be incorporated into the model as it is.

### 3.3. Female model

The difference between the male and female income distribution in the U.S. is a well-established fact. We do not consider the reasons behind the catastrophic disparity, but have to stress that the work capabilities used in the model to predict personal incomes are likely the same for men and women. From technical point of view, this difference allows to improve the model and introduce new features which were not observed in the overall PID and its aggregates and derivatives. The updated model successfully meets a number challenges.

There are features in the females' curves in Section 2, which likely manifest some changes in the parameters considered in the original setting as constant. The convergence trend of the male and female PIDs observed since the earlier 1960s suggests that the size of work instruments available for women has been growing as a portion of standard sizes. It does not reach the level of the males' instruments, however. So far, we used fixed relative sizes of work instruments. So, our model has to include a new option allowing the size change according to some predefined time or real GDP function.

It is reasonable to start with an approximate estimate of FL, which fits best the observed features for different years. We have calculated a number of models with FL changing from 0.2 to 1.0 with a 0.05 step and all other parameters as in the original model. As we know from Section 2, the most sensitive part in the mean income dependence on age is the initial segment. Figure 28 depicts the measured and predicted mean income curves for 1962 and 1977 as obtained with FL=0.45. This FL value is the best to describe the dynamics of mean income growth in the youngest population and also demonstrates the feature of constant mean income before the critical age. The predicted curves start to fall long before the critical age, however. But $T_c$ is controlled by a different dependence on real GDP per capita and does not influence the initial growth. So, the estimate of FL=0.45 is not compromised by the deviation of the measured and predicted critical ages, which we have address next.

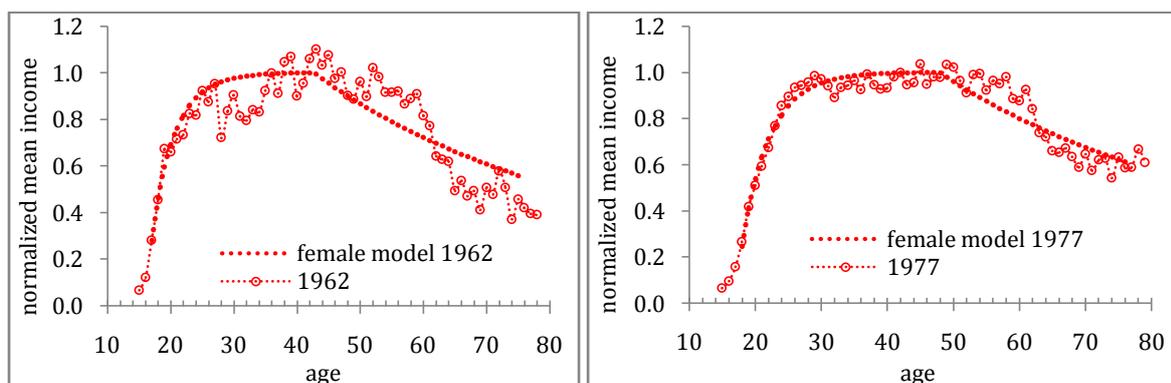

Figure 28. The observed and predicted mean income for 1962 (left panel) and 1977 (right panel). In both models FL=0.45.

In Figure 29 we present similar curves for 1996 and 2011, but for FL=0.65. The measured curves are accurately approximated by the model between the start point and the critical age. The fall is also well predicted but there are some deviations in 1996, which could have some connection to retirement, as discussed in the previous Subsection. Comparing the curves in Figures 28 and 29 one can conclude that FL has to grow from 0.45 in 1962 to 0.65 in 2011 in order to fit the rate and duration of the initial growth.

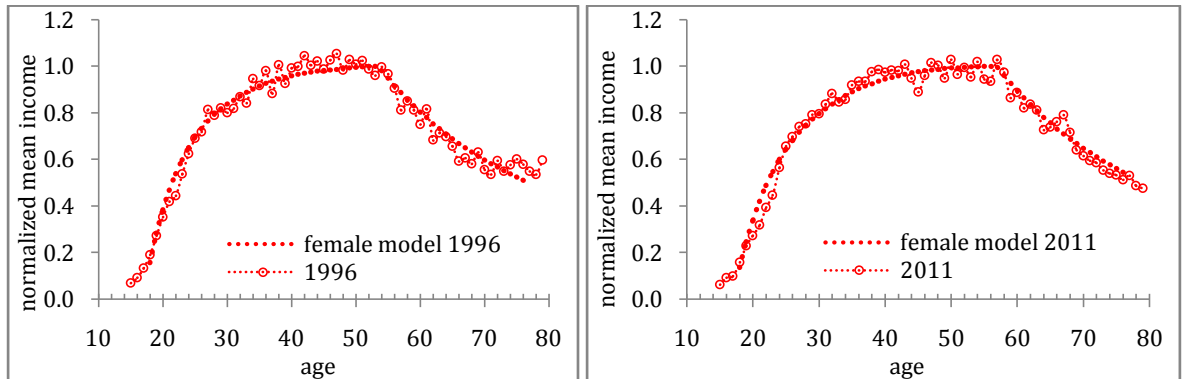

Figure 29. The observed and predicted mean income for 1996 (left panel) and 2011 (right panel). In both models FL=0.65. Notice the fit between the observed and measured critical ages.

The simplest way to match the critical age observed between 1962 and 1977 is to rise the initial value $T_c(1929)$. Then the predicted value in 1962 and 1977 has to change accordingly as the square root of real GDP per capita. Figure 30 displays the modified mean income curves (green dotted lines) for the same years as in Figure 28 and 29, which now fit observations in 1962 and 1977 before and beyond the critical age. For comparison, we have also drawn the curves from the original model (red dotted lines). The change in $T_c$ is not justified by any reasonable relationship, however. In addition, the curves predicted for 1996 and 2011 with the higher initial value of $T_c$ do not fit observations neither in the initial segment not in the critical age. The shelf exists in all models, but its length is not well predicted. Considering the large number of contradictions we met when modelling the female mean income it is necessary to extend our original model to match all observed changes in a consistent way.

A lower FL in the earlier years provides an extended "no-change" period before the critical age. The difference between the observed and predicted critical ages in Figure 28 suggests that the concept of critical age belongs to the super-critical distribution rather than to the low-middle incomes predicted by the model. Moreover, the critical age for women observed between 1962 and the earlier 1980s is almost constant and close to the retirement age. It is reasonable to suggest that there are two critical ages – one for the super-critical distribution, $T_c$, and another for the sub-critical distribution, $T_S$. The former variable depends on GDP. The latter one – is practically constant and may correlate with retirement age. It is easy to model the retirement effect by an exponential fall in the portion of people in the labour force [Munell, 2011]. Using the basic concept of (18) one can write the following equation:

$$\eta = -\ln B / (T_B - T_S) \tag{23}$$

where $\eta$ is the exponential index, $B$ is the constant relative level of income rate at age $T_B > T_S$. Both constants in (23) have to be estimated from data. Then the evolution of sub-critical incomes above the retirement age is described by (23) and the evolution of super-critical incomes is defined by (18). When a super-critical income falls below the Pareto threshold it starts to follow (23).

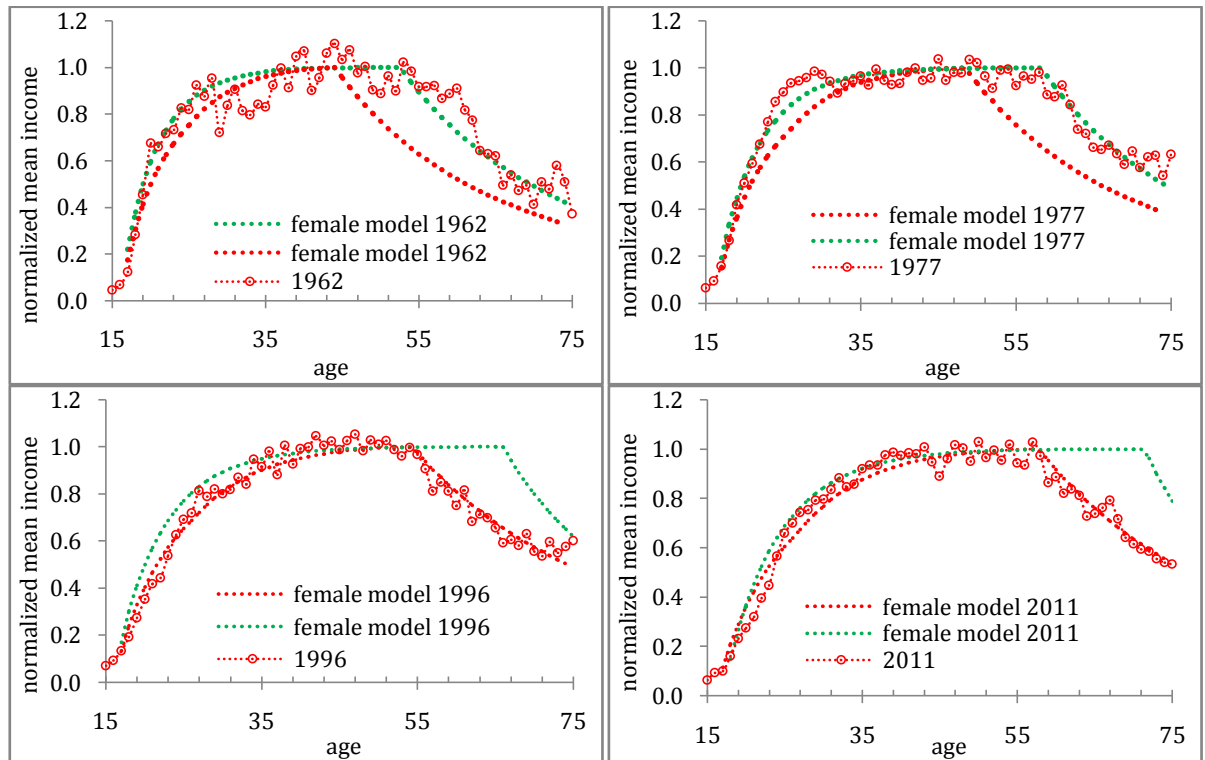

Figure 30. The observed and predicted mean income as a function of age for selected years between 1962 and 2011. Two models are shown: standard model with FL=0.45 and $T_c$=19.07 years in 1929 (red dotted line); standard model with FL=0.45 and $T_c$=26.0 years in 1929 (green dotted line).

The effect of $T_S$ in the overall distribution was masked by income dominance of male population, data roughness, and by the fact that only the most recent period was modelled. As a result, $T_S$ is missing from the original model. The use of the detailed IPUMS income microdata available from 1962 has revealed this effect for both genders. Figure 26 indicates that this effect cannot be neglected even in the males PIDS and we consider it later on.

Having the new concept of critical age, $T_S$, for sub-critical incomes we can model now the lengthy shelf observed in the mean income curves. Three parameters in (23) have to be fixed in the model: $T_S$, $T_B$, and $B$. The age of retirement may vary by a few years over the whole modelled period between 1962 and 2011 as we see in the measured mean income curves. Without loss of generality, from the lengthy period of income measurements, from the estimated age of women's retirement, as well as from the results of extensive modelling we fix $T_s$ to 43 years of work experience or 58 years of age. This is likely the start of retirement process, which is described by an exponential fall in the portion of people in labour force. The estimated age of retirement is defined by the portion of 50% in the labour force [Munell, 2011]. To match observations we vary $A$ and $B$ in (18) and (23) and obtain the best fit estimate by a simple regression procedure in the age ranging from 15 to 70 years. Larger deviations observed beyond 70 years of age are likely related to poor measurements and should be neglected as noise.

From Figure 30, we have estimated the overall increase in FL from 1962 to 2011. From the IPUMS data, one cannot estimate the exact form of functional dependence of women's $L_j$ on time or real GDP per capita before 1962. It is reasonable to choose 1960 as the start year and to reduce all initial values used in the original model starting in 1930 to 1960. They are as follows: $\tau_0$=1960, $T_c$=25.92 years, and $\alpha$= 0.0795, $Y(1960)$=1. For the dependence of FL on time, we suggest the simplest linear time growth between 1962 and 2011. It makes a 0.004 annual step in FL starting from 0.45 in 1960. We understand that the actual FL dependence on time may deviate from linear approximation, but the accuracy of the IPUMS data is not good enough to estimate exact FL for each year between 1962 and 2011.

There is one model parameter pending adjustment – the Pareto threshold for women, which likely increases over time. As we discussed in Section 3.1, the Pareto threshold may have a direct link to the size of work instrument. Interestingly, the reduction factor FL for women can be best estimated from the rate of growth at the initial stage before 30 years of age. The Pareto distribution is most prominent for the ages above 30 and below 60. But they are connected by some immanent bond, which manifests itself in a slightly different way for women and men.

Observations in Figures 9 and 22 suggest that the power law approximation works from different incomes for males and females. For example, in 1962 the Pareto threshold is $9,000 for males and $6,000 for females. If to consider the males' threshold as related to normal sizes of work instruments, then the females' threshold should be reduced by a factor of 2/3. It makes $M^P$=0.29 in 1960 and then this threshold should increase to the level close 0.43 in 2011. In order to match the linear growth in FL, we suggest a similar time function for $M^P$ for women. The initial value $M^P(1960)=0.29$ and then it rises at a rate of 0.002 per year, making 0.39 in 2010. Then the growth in $M^P$ and FL are synchronized.

There are several novelties proposed in Section 3 to accommodate a few specific income distribution features observed for females. First of all, we have found that the processes in the sub-critical and super-critical income zones have different critical ages, $T_c$ and $T_S$, and indexes of exponential fall - relationships (18) and (23). Essentially, the high-income dynamic processes are decoupled from those in the low-middle income zone. The only connection between them is the exchange of people reaching the Pareto threshold from below and above. The joint distribution of incomes driven by different processes in two income zones is able to describe all features observed in the mean income curves for females.

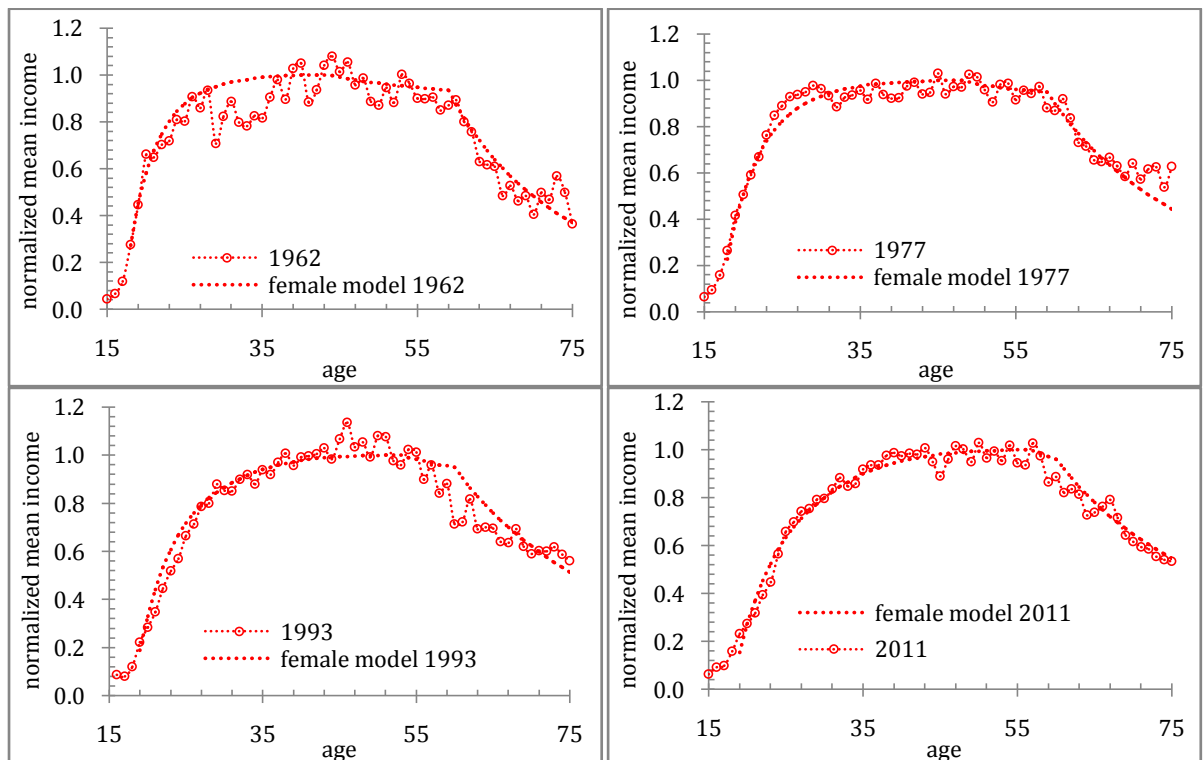

Figure 31. The observed and predicted mean income as a function of age for selected years between 1962 and 2011.

The multivariable matching process results in the best fit model over all involved parameters. Figure 31 depicts the observed and predicted mean income as a function of age for four years between 1962 and 2011. The fit of the initial segments is the same as in Figure 28 and 29 because the reduction factor FL for the size of instruments to earn money changes from 0.45 in 1960 to 0.65 in 2011. The shelf modelled in Figure 30 by different critical times $T_c$ now consists of two segments as related to the critical ages in the sub-critical ($T_S$) and super-critical ($T_c$) zones. Since $T_S$ is constant, the duration of the second leg, i.e. $T_S$-$T_c$, has been decreasing with time while the duration of the first leg has been increasing. The second leg of the shelf has a length of a few years in 2011. Following this trend, the second leg will disappear in the future when $T_c$>$T_S$. The U.S. political and economic authorities may increase the age of retirement, however.

The overall fit between the observed and predicted mean income curves in Figure 31 is much better than that obtained by the original model. In 1993, we notice a slightly lower agreement between the observed and measured mean income during the second leg of the shelf. This is likely the result of the linear approximation of FL and $M^P$. In reality, these parameters are different from those predicted by linear time function. It is worth noting that this discrepancy disappears in 2011 because of the overall convergence between the males and females characteristics. In a few decades, the male and female PIDs should be identical. Meanwhile, it is instructive to apply the upgraded model to males.

### 3.4. **Male model refined**

The relative sizes of work instruments $L_j$ and the Pareto threshold $M^P$ for males do not change with time as well as the age of retirement $T_S$. This fact facilitates the use of the model refined in Section 3.3. We have to adjust only the levels *A* and *B* in (18) and (23) in order to obtain the best fit model. Figure 32 depicts the measured and predicted mean income curves for six years between 1962 and 2011. The overall fit is better than that achieved in Section 3.2 (see Figure 26). The fit below the critical age $T_c$ is the same, however, since it is controlled by the parameters, which were not changed in the upgraded model.

There are two distinct segments beyond the critical age for the super-critical incomes. First segment spans the ages between $T_c$ and $T_S$, and thus, shortens with time. Because of the larger portion of males above the Pareto threshold the first segment demonstrates a clear fall, which is not well seen in Figure 31. The slope of this segment also increases with time since the exponent index $\tilde{\gamma}$ in (18) depends of $T_c(\tau)$. The second segment of the mean income curves is driven by to processes of exponential fall. The highest incomes continue to fall as in the first segment and the sub-critical incomes start their fall beyond the age of retirement. The joint effect of two processes is expressed by an expedite drop beyond $T_S$.

The first segment related to the super-critical incomes is predicted with an excellent accuracy for all presented years as well as the trajectories before $T_c$. This is an important verification of the model predictive power. The changing slope and length of the pre-critical stage ($t<T_c$) and same features of the first segment - all change over time as defined by their dependence on one external variable – real GDP per capita. The predicted curves not only fit the measured ones for the selected years. They demonstrate synchronized change with time (real GDP) – the ultimate demand of any dynamic model in physics. Hence, we conclude that our microeconomic model is a physical one rather than economic.

The prediction of the second segment is complicated by a few shortcomings. Firstly, the age of retirement changes with time depending not only on legislation but also on economic situation. After the last financial and economic crisis, many discouraged people have to leave the labour force. The rate of participation in labour force has fallen from 66.4% in 2007 to 62.4% in September 2015. (During the same period the rate of unemployment has changed from 4.6% to 5.1%, with an extremely high rate in-between, however.) As a result the age of retirement may drop. Secondly, the accuracy of income measurements is significantly lower for the eldest population because of the overall under-representation. Thirdly, the population pyramid for the elderly fluctuates as a consequence of WWII and the post-war baby boom. Our model is able adjust all these aspects and predict the second segment better than its current version. This would be non-physical amendments, however, with are not the priority of quantitative modelling.

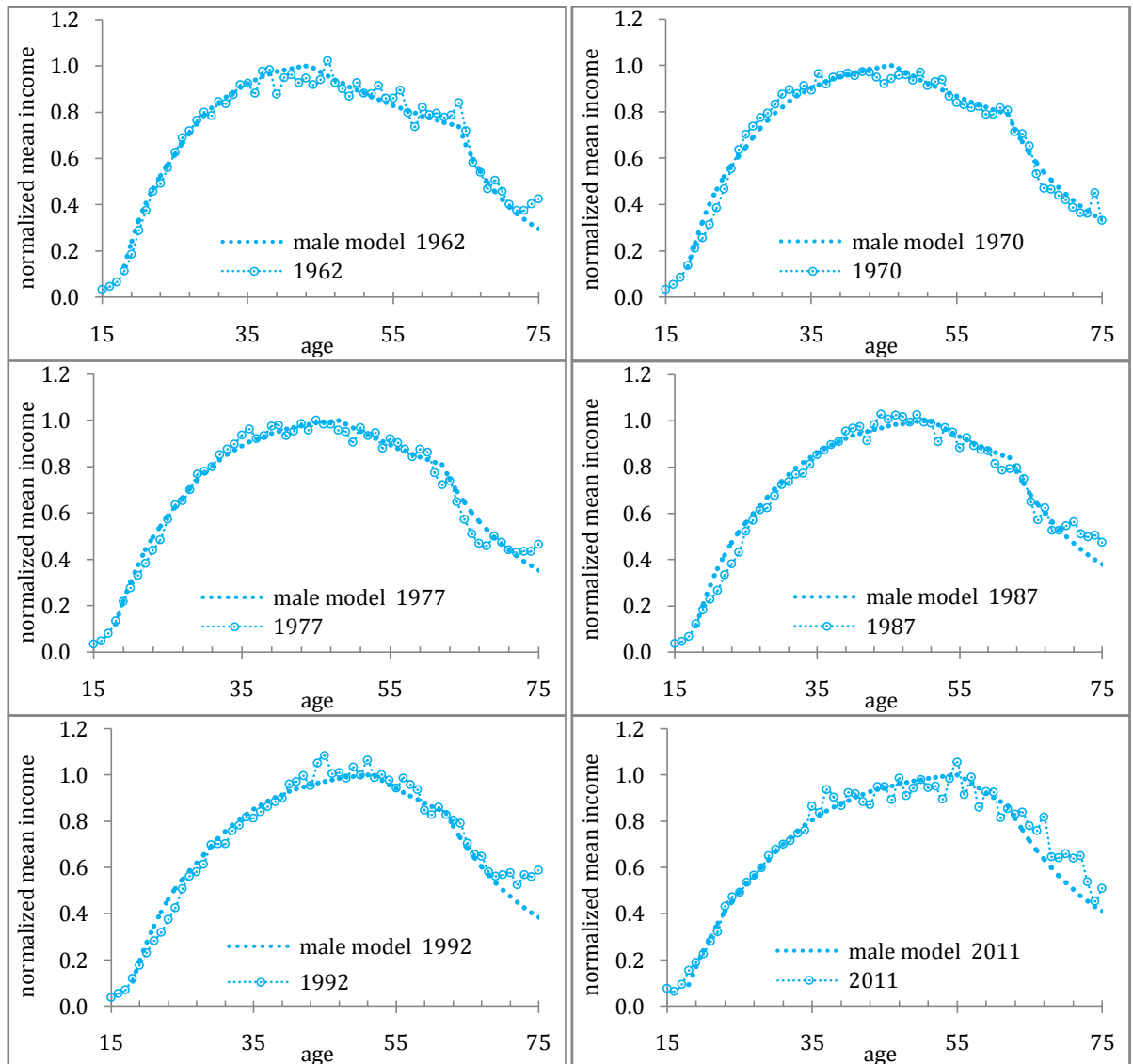

Figure 32. The observed and predicted mean income curves for male population as a function of age.

## Discussion

Our model stems from extensive physical intuition and is supported by direct comparison of income observations with closed-form solutions of simple differential equations describing fundamental physical processes. The set of equations describing the growth and fall of incomes is fully borrowed from physics, with the empirically estimated constant of dissipation and the distribution of sizes of personal capabilities and instruments. The transition to the power law distribution of the highest incomes is also a physical process, which can be qualitatively described by the concept of self-organized criticality. In that sense, the dynamics of the highest incomes is not governed by simple physical relationships – the power law distribution is a purely statistical description rather than a solution of a system of differential equations. Nevertheless, all properties in the super-critical regime are defined by two parameters – the number of people above the Pareto threshold and the power law index. The former parameter is exactly predicted by our model as a function of time and age for males and females. The index has to be empirically estimated, as in other physical cases like for the slopes of earthquake recurrence curves in various seismic regions [Kitov *et al.*, 2011]. Overall, the system of personal income distribution is fully and accurately described by physical equations. It is a physical system.

We introduced the microeconomic model of personal income distribution a decade ago and used the CPS historical data to calibrate all defining parameters. The March CPS reports aggregate incomes in five-year age cells since 1993. Ten-year cells are used between 1947 and 1992, with sporadic appearance of shorter cells for the youngest population. The income data and the U.S. age pyramids between 1947 and 2011 published by the U.S. Census Bureau were used as they are without gender separation. The IPUMS income microdata not only make it possible to distinguish between males and females but also provide various estimates in one-year cells. In this study, we use the advantage of income microdata and model two specific age-dependent features of personal income distribution: mean income and portion of people in the Pareto distribution. These two features are most sensitive to the influence of time and age. A correct income distribution model must accurately describe the dynamics of secular and age-dependent changes observed in actual data. Any model not predicting the dynamics of actual changes should be disregarded. Our model successfully predicts all principal changes in both features observed between 1962 and 2014 for males and females separately.

The difference in income dynamics demonstrated by two genders represents enormous challenge for quantitative modelling. A model unifying incompatible (at first glance) results for two genders has to be parsimonious and include only parameters common for both cases. The dynamic discrepancy between male and female incomes has to be explained only by values of defining parameters: constants and variables controlled by exogenous measurable forces represented by continuous time series. The evolution of gender-dependent income features together with all changes in the difference between them should be driven by the same driving forces. In our model, the only force moving personal income distribution along predefined trajectories is real economic growth as expressed by GDP per capita calculated for working age population.

Dynamic behavior of the difference in income distribution between males and females requires a special approach in quantitative models of income distribution. The original version of KKM made no difference between men and women. Here, we extend the KKM by introducing two independent populations with different features of income distribution as reported by the CPS and IPUMS. Since gender divides U.S. population in approximately equal proportions over time and age the gender-related income effects do improve the KKM predictive power upon the original version. In other words, females have sizeable contribution to the total income. The next step to a more precise model might be the introduction of race differences of income distribution. The income difference between white males and black females is much more dramatic than income difference between two sexes considered in this study. This is a real challenge to our income distribution model.

All in all, we have demonstrated in this paper that the refined KKM accurately explains a number of common and gender-specific features. The principal finding of this study is that female population in the U.S. has the same distribution of the capability to earn money (notation similar but not equivalent to human capital) and consistently lower sizes of work instruments (work capital) compared to those for men. The income gap between women and men has been closing since 1960 and currently an average female has work capital making 65% of that available for an average man. It was only 45% in the 1960s. Considering the same capability to earn money for females, one can conclude that the relatively lower work capitals (*e.g.,* job positions, assets, …) are controlled by external force. A fair distribution has not been achieved yet. It will likely take decades.

The relatively lower instrument sizes available for females make the proportion of female above the Pareto threshold lower. In turn, this effect lowers the mean income for the same age since a relatively lower number of rich females occurs in all age groups. The lack of rich women is partially compensated by the effect of lowered Pareto threshold for females, which is most prominent in the 1960s and 1970s. The coherent increase in the instrument size and Pareto threshold for women has been incorporated into our model. As a result, the model accurately predicts the early growth trajectory, which is most sensitive to the size of work instrument, and the number of females above their own Pareto threshold. As in the original model, both parameters increase with time as the square root of real GDP per capita. For women, we have introduced a specific option as revealed from observations - the relative instrument size and the Pareto threshold both follow linear time trends with different slopes.

The female mean income shows a very specific feature – it is practically constant during an extended period spanning the ages between ~30 and ~60. In our model, this feature results from the fast growth of all personal incomes to their peak values, which are then retained at the same level. The expedite rise in all incomes is induced by the lowered sizes of work instruments available for women. In turn, the lower instruments do not allow personal incomes to reach the Pareto threshold and there are almost no rich women by male standards in the 1960s and 1970s. Therefore, the disparity in work capitals affects the low-middle incomes and higher incomes together. Such a shelf is absent in the overall mean income curve because of larger instrument sizes available for males.

The shelf in the females' mean income curves has also revealed the difference between critical times for the low-middle (in physical notation - sub-critical) and high (super-critical) incomes, the latter governed by the Pareto distribution. Equation (13) describing the sub–critical regime is valid from the start of work experience to the age of retirement. Then incomes fall along an exponential trajectory described by equation (17). The actual age of retirement varies in a narrow band between ~60 and ~65 years and is embedded into the model as constant. The fall is described by an exponential function with a negative index. This is a new feature of the upgraded model. In the original model, the critical age, $T_c$, was the same for low-middle and high incomes. The input of rich men in the overall PID masked the presence of the low-middle income critical age. Instructively, the mean income measured for males supports the existence of two critical ages.

The refined model does include several new features not compromising the underlying physical concept of saturation growth and the transition from sub-critical to super-critical regime of income distribution. The extended version of the original model accurately predicts the PID evolution for males and females in the U.S. from 1962 to 2014, *i.e.* where the IPUMS data are available. Since the GDP estimates are available from the U.S. Bureau of Economic Analysis since 1929 we start our model in 1930 for males. Actually, the model spans the period since 1870, *i.e.* the year when started their work people who reached the age of 75 in 1930. For females, the start year is shifted to 1960 because of changing relative size of work instrument and Pareto threshold.

Forced deprivation of higher job positions (work capital) is the cause of the observed long term income inequality between male and female in the U.S. It is not only unjust to women but has a negative effect on real economic growth. The replacement of highly capable women with less capable men results in lower total income, which is an equivalent to real GDP. Women have been catching up since the 1960s and that improves the performance of the U.S. economy. It will take decades, however, to full income equality between genders. The problem of race income disparity will take longer time to full resolution, however.